 \documentclass[final,3p,times]{elsarticle}
 \makeatletter
 \def\ps@pprintTitle{%
 	\let\@oddhead\@empty
 	\let\@evenhead\@empty
 	\def\@oddfoot{\reset@font\hfil\thepage\hfil}
 	\let\@evenfoot\@oddfoot
 }
 \makeatother

\usepackage{lineno,hyperref}

\usepackage{amsmath}
\usepackage{graphicx}
\usepackage{amssymb} 
\usepackage[]{algorithm2e}
\usepackage{breqn}

\usepackage{subcaption}

\usepackage{xspace}
\newcommand*{\ie}{i.e.\@\xspace}
\newcommand*{\eg}{e.g.\@\xspace}
\newcommand*{\fig}{Fig.\@\xspace}
\newcommand*{\eq}{eq.\@\xspace}

\newcommand*{\rhs}{r.h.s.\@\xspace}

\usepackage{braket}

\newcommand{\ex}{\text{e}}

\usepackage{xifthen}
\newcommand*\diff{\mathrm{d}} 
\newcommand*\ldiff[2][]{ \ifthenelse{\isempty{#1}}{ \diff #2}{\diff^#1#2} \,} 
\let\limitint\int 
\renewcommand{\int}{\limitint \!} 

\usepackage{xcolor}
\newcounter{bla}

\begin{document}

\begin{frontmatter}



\title{\textit{TimeEvolver}:\\ \mbox{A Program for Time Evolution With Improved Error Bound}}


\author[bguaddress]{Marco Michel}
\ead{michelma@post.bgu.ac.il}

\author[epfladdress]{Sebastian Zell}
\ead{sebastian.zell@epfl.ch}

\address[bguaddress]{Department of Physics, Ben-Gurion University of the Negev, Beer-Sheva 84105, Israel}
\address[epfladdress]{Institute of Physics, Laboratory for Particle Physics and Cosmology,
	\'Ecole Polytechnique F\'ed\'erale de Lausanne, CH-1015 Lausanne, Switzerland}

\begin{abstract}
	We present \textit{TimeEvolver}, a program for computing time evolution in a generic quantum system. It relies on well-known Krylov subspace techniques to tackle the problem of multiplying the exponential of a large sparse matrix $i H$, where $H$ is the Hamiltonian, with an initial vector $v$. The fact that $H$ is Hermitian makes it possible to provide an easily computable bound on the accuracy of the Krylov approximation. Apart from effects of numerical roundoff, the resulting a posteriori error bound is rigorous, which represents a crucial novelty as compared to existing software packages such as \textit{Expokit} \cite{Sidje1998}. On a standard notebook,	\textit{TimeEvolver} allows to compute time evolution with adjustable precision in Hilbert spaces of dimension greater than $10^6$. Additionally, we provide routines for deriving the matrix $H$ from a more abstract representation of the Hamiltonian operator.
\end{abstract} 


\end{frontmatter}



{\bf\noindent Program summary}\\ 

\begin{small}
\noindent
{\em Program Title:} TimeEvolver    \\
{\em Link to published article in Computer Physics Communications:}   \href{https://doi.org/10.1016/j.cpc.2022.108374}{doi.org/10.1016/j.cpc.2022.108374} \\                                    
{\em CPC Library link to program files:} \href{https://doi.org/10.17632/vvwvng9w36.1}{doi.org/10.17632/vvwvng9w36.1} \\
{\em Developer's repository link:} \href{https://github.com/marco-michel/TimeEvolver}{github.com/marco-michel/TimeEvolver} \\
{\em Code Ocean capsule:} \href{https://codeocean.com/capsule/8431379}{codeocean.com/capsule/8431379}\\
{\em Licensing provisions:} MIT\\
{\em Programming language:} C++                                   \\
{\em Supplementary material:} An example which demonstrates the computation of time evolution in a concrete physical system \\
{\em Nature of problem:} Computing time evolution in a generic physical quantum system can be reduced to the numerical task of calculating $\exp(-iHt) v$. Here $H$ is the Hamiltonian matrix, which is large and sparse, $i$ corresponds to the imaginary unit, $t$ denotes time and the vector $v$ represents the initial state. A program is needed to  perform this computation efficiently. Since the use of approximation methods is unavoidable, it is important to quantify as rigorously as possible the resulting error. Moreover, in order to facilitate the application to various problems in physics, additional functionalities are needed, in particular for forming the Hamiltonian matrix from a more abstract representation of the Hamiltonian operator.\\
{\em Solution method:} The program employs known Krylov subspace methods for calculation the exponential of the large sparse matrix $(-iHt)$ times the vector $v$. The Arnoldi algorithm is used to form the Krylov subspace and exponentiation of the resulting small matrix is achieved by diagonalization. The fact that $(-iHt)$ is anti-Hermitian makes it possible to calculate the error of the Krylov approximation in terms of an easily-computable integral formula. This allows to choose a maximal size of the time step, after which the method is restarted and a new Krylov subspace is formed, while respecting an adjustable error bound. It is rigorous up to inaccuracies of a one-dimensional numerical integral and effects of finite machine precision, for which we also give an estimate. All linear algebra operations are performed with the Intel\/\textsuperscript{\circledR} Math Kernel Library and Boost is used for numerical integration. The methods for deriving the Hamiltonian matrix rely on a hashtable representation of Hilbert space.\\


\end{small}

\tableofcontents

\section{Introduction}

\subsection{Time Evolution in a Generic Physical Quantum System}

We consider a generic physical quantum system. At a given time $t$, all information about it can be encoded in a state vector $v(t)$, \ie knowledge of $v(t)$ allows to predict the outcome of any conceivable measurement. All states that a system can possibly assume are contained in a Hilbert space $\mathcal{H}$. According to our understanding of physics, our Universe is deterministic in the following sense. The state $v(0)\in \mathcal{H}$ at some initial time uniquely determines the states $v(t)\in \mathcal{H}$ at all other times. A central task in physics is to infer $v(t)$ from the knowledge of $v(0)$. 

In principle, this is very easy. The only additional ingredient that one needs is a Hamiltonian $H$, which is a self-adjoint and time-independent operator on the Hilbert space $\mathcal{H}$.\footnote
{Time-independence of the Hamiltonian ensures that energy is conserved. Therefore, any fundamental Hamiltonian must not depend on time. Nevertheless, it is sometimes possible to describe certain subsystems of a bigger physical system in terms of an effective time-dependent Hamiltonian. We shall not consider this case in the present paper.}
Given $v(0)$ and $H$, it follows from the Schr\"odinger equation that the state at any other point in time $t$ can be computed as
\begin{equation} \label{vOfT}
	v(t) = \ex^{-i H t} v(0) \,.
\end{equation}
As said the vector $v(t)$ represents a full solution, \ie it completely determines the system. Thus, finding techniques for calculating the \rhs of \eq \eqref{vOfT} bears great relevance for all physical systems.

In some situations, analytic methods and approximations are available for performing this computation. For a generic system, however, the only way to compute the \rhs of \eq \eqref{vOfT} is to employ a numerical approach. Then $v(0)$ is a vector with $d$ complex entries and $H$ is a Hermitian $d\,$x$\,d$ matrix, where $d$ is the dimension of the Hilbert space $\mathcal{H}$. Each column of $H$ encodes to which states a given basis state can transit. Typically, this number is much smaller than $d$, \ie $H$ is a sparse matrix. Thus, one has to compute the product of an exponentiated anti-Hermitian sparse matrix and a vector. Exaggerating slightly, one could say that solving any physical system reduces to this numerical task.

\subsection{Krylov Subspace Methods}
Of course, a plethora of approaches exists to exponentiate a matrix. But the special nature of the problem \eqref{vOfT} -- especially the fact that in the end knowledge of the full matrix $\ex^{-i H t}$ is not required -- allows the use of techniques that are significantly more efficient than generic routines for matrix exponentiation. Particularly important are Krylov subspace methods, which were first used in 1986 for the purpose of numerical time evolution \cite{Park1986}. 
We shall brief\/ly sketch them now, before we review them in more detail in section \ref{sec:krylovIntro}. 

The $m$-dimensional Krylov subspace $\mathcal{K}_m$ is defined as 
\begin{equation} \label{krylovSpace}
	\mathcal{K}_m := \text{span}\left\{v(0), Hv(0), \ldots, (H)^{m-1}v(0)\right\} \,,
\end{equation}
where $m\ll d$. (Typically, one has $m \lesssim 100$.) The key idea is to project the large sparse matrix $H$ (in our case the Hamiltonian) on the small subspace $\mathcal{K}_m$:
\begin{equation} \label{projection}
	H_m := H\vert_{\mathcal{K}_m} \,.
\end{equation}
Since $H_m$ represents a mapping from $\mathcal{K}_m$ to $\mathcal{K}_m$, it corresponds to a small $m\,$x$\,m$ matrix. Now the crucial step is to effectively replace 
\begin{equation} \label{approximationSymbolic}
	\ex^{-i H t} \quad \rightarrow \quad \ex^{-i H_m t} \,,
\end{equation}
\ie to use instead of the exact result $v(t)=\ex^{-i H t}v(0)$ the Krylov approximation 
	\begin{equation} \label{KrylovResultSymbolic}
	\tilde{v}(t)= V_m \ex^{-i H_m t} V_m^T v(0) \;,
	\end{equation} 
where $V_m$ stands for the embedding of $\mathcal{K}_m$ in the full Hilbert space $\mathcal{H}$. In doing so, we introduce the error
\begin{equation} \label{error}
	\text{err}(t) := v(t)-\tilde{v}(t) \,.
\end{equation}
 Once matrix exponentiation is restricted to the small subspace $\mathcal{K}_m$ as shown in \eq \eqref{KrylovResultSymbolic}, it can be performed very fast using standard methods. In the full Hilbert space $\mathcal{H}$, only matrix-vector multiplications, as displayed in \eq \eqref{krylovSpace}, need to be carried out. Thus, the effectiveness of Krylov subspace methods is due the fact that no matrix-matrix multiplication has to be performed in the large Hilbert space $\mathcal{H}$. 

We can estimate how large the advantage of Krylov subspace methods is. Clearly, a necessary requirement for the feasibility of the calculation is that all intermediate results can be stored in (the memory of) a computer. It turns out that this condition, and not runtime, indeed is the limiting factor in many practical applications. If one uses standard matrix exponentiation, the biggest intermediate object is the matrix $\ex^{-i H t}$, which unlike $H$ is no longer sparse. So the required storage capacity scales as the number of entries, \ie $d^2$. In contrast, the largest entities that are needed for the Krylov method are the basis elements of the Krylov space \eqref{krylovSpace}, so the required memory scales as $m\, d$. In practice, this leads to a gain in Hilbert space size of several orders of magnitude.

The Krylov subspace \eqref{krylovSpace} is specifically adapted to the initial state $v(0)$. Therefore, one can expect that for short enough times $t$, the error \eqref{error} is small whereas it grows large once the state $v(t)$ (as determined without any approximation) deviates sufficiently from $v(0)$. Thus, it becomes necessary to restart the procedure after a certain time interval, \ie to take $\tilde{v}(t)$, computed by the Krylov method, as initial state for a new Krylov subspace. With such a time stepping procedure in place, it has become evident since the earliest numerical studies that in many circumstances the Krylov method works surprisingly well already for values of $m$ as small as $10$ (see \eg \cite{Park1986, Gallopoulos1989, Saad1992, Gallopoulos1992}).

At the same time, these and subsequent investigations have lead to significant conceptual progress in understanding how these Krylov subspace techniques work.
A possibly incomplete list of relevant studies includes \cite{Gallopoulos1989, Saad1992, Gallopoulos1992, Sidje1994, Philippe1995,Druskin1995, Stewart1996, Hochbruck1997, Celledoni1997, Druskin1998, Hochbruck1998}, where it is important to mention that these publications have a wider scope than the present paper in that they deal with the exponentiation of a generic sparse matrix.
In the works cited above, a particularly relevant object of study is to provide bounds on the error \eqref{error}. In general, one can distinguish between a priori and a posteriori error bounds. The first ones can be evaluated without actually performing the Krylov technique while the second ones rely on the result of the Krylov method. Correspondingly, a priori error bounds tend to be more general (and in particular independent of the initial state $v(0)$) whereas a posteriori error bounds are often sharper. 

However, both types of rigorous error bounds typically suffer from the problem that the function bounding the error is difficult to evaluate in practice. Therefore, a posteriori error estimates have been introduced in \cite{Park1986, Saad1992}. They play a twofold role in numerical implementations of Krylov subspace methods. First, they are used to determine the time interval after which restarting of the Krylov procedure becomes necessary. Secondly, they can be used to estimate the error of the final result. 
Obviously, using error estimates instead of error bounds leads to the problem that one can in general not be sure if the outcome of the Krylov method is indeed close to the true result.

\subsection{Practical Computation of Time Evolution}
As we have discussed, Krylov subspace methods represent an efficient approximation scheme for calculating the product of an exponentiated sparse matrix and a vector. Such a problem appears in a wide range of context. In particular, it is relevant for computing time evolution in a generic quantum system, as displayed in \eq \eqref{vOfT}. In this case, the sparse matrix $-i H t$, which is exponentiated, is anti-Hermitian. From the perspective of Krylov methods, this represents a special case, but in a physical context this encompasses the most general quantum system.

In order to actually compute the result of time evolution in a given physical system, a numerical implementation of Krylov subspace methods is required. This crucial tool was provided by Sidje in 1998 when he published the software package \textit{Expokit} \cite{Sidje1998}.\footnote
{In particular, it is based on the theoretical investigations \cite{Gallopoulos1989, Saad1992, Gallopoulos1992, Sidje1994,Philippe1995}.}
Among other routines for matrix exponentiation, this program provides Krylov subspace methods for generic sparse matrices. Since certain simplifications occur for (anti-)Hermitian matrices, \textit{Expokit} contains functions that are specific to this special case, which is extremely important in the context of physics. 
In line with our previous discussions, \textit{Expokit} does not employ a rigorous error bound, but it relies on the error estimate proposed in \cite{Saad1992} and subsequent refinements developed in \cite{Sidje1994, Philippe1995}.

The lack of a rigorous error bound represents a serious drawback for any practical application. If the problem at hand is such that one can compute the result by means of a different method, then one can of course check the validity of the error estimates. However, in such a case there is no need to employ Krylov subspace techniques in the first place. This is to say that Krylov approximation methods are only interesting if there is no other way of solving the problem. But in this case it is impossible to make sure that the error estimates are still viable. Therefore, the numerical results obtained from Krylov subspace methods would become significantly more meaningful if a more precise statement about the error, ideally a rigorous error bound, were available.

Since the release of \textit{Expokit}, much progress has been made in the study of a posteriori error bounds and corresponding criteria for choosing the size of the time step \cite{Celledoni1997, Druskin1998, Hochbruck1998, Lubich2008, Botchev2013, Ye2013, Jia2015, Wang2016, Jawecki2018}. To our knowledge, Lubich was the first one to point out that significant simplifications occur in the physically relevant case in which an anti-Hermitian matrix is exponentiated: The fact that the matrix norm of  $\ex^{-i H t}$ is $1$ leads to the existence of an a posteriori error bound, which can be represented as an integral over quantities that are already known within the Krylov approximation \cite{Lubich2008}. In this way, one obtains a formula for computing the error bound that is both rigorous and easy to compute in practice. 
Of course, it is important to remark that a rigorous formula for bounding the error of the Krylov approximation still does not lead to a fully rigorous statement about the accuracy of a numerical result. The reason is that a finite machine precision inevitably causes additional roundoff errors. Nevertheless, reducing the uncertainty in the error to effects of finite machine precision certainly represents an improvement.

The goal of the present work is to incorporate this knowledge, and in particular the error bound proposed in \cite{Lubich2008}, in a concrete numerical implementation of Krylov subspace methods. To this end, we have developed the software package \textit{TimeEvolver}, which is able to compute numerical time evolution while obeying an adjustable error bound. It is rigorous up to effects of roundoff errors and a negligible uncertainty due to a finite step size in the evaluation of the error integral. Moreover, we provide an estimate for the numerical roundoff error. Since we specialize to anti-Hermitian matrices from the outset, our program is considerably less general than \textit{Expokit}. However, it is sufficient for the most general application of quantum mechanics, namely time evolving a generic physical system. In turn, we have aimed at making our program easy to use and to modify. Correspondingly, \textit{TimeEvolver} is open-access and based on free software. In this way, we hope to make Krylov subspace methods easily accessible for physicists from a wide range of fields.

The outline of this paper is as follows. Section \ref{sec:method} starts with a more detailed review of known Krylov methods. We pay particular attention to a posteriori error bounds and also present our estimate for the numerical rounding error. Subsequently, we describe our concrete algorithm.
Section \eqref{sec:numberOperators} discusses the practical problem of deriving the Hamiltonian matrix from a more abstract Hamiltonian operator. We specialize to the widely-used situation in which the basis of Hilbert space is given by number eigenstates.
Section \ref{sec:implementation} contains a description of \textit{TimeEvolver}, \ie the concrete implementation of our approach in C++. In section \ref{sec:performance}, we apply our program to an exemplary physical system and study its performance. We summarize in section \ref{sec:summary}. Throughout, we try to be self-contained and not assume specific prior knowledge. In this way, we hope to make our presentation also understandable to a physicist without previous experience in the application of Krylov subspace techniques.

\section{The Method}
\label{sec:method}

\subsection{Krylov Subspace}
\label{sec:krylovIntro} 
Following \cite{Gallopoulos1989, Saad1992, Gallopoulos1992}, we shall review known Krylov subspace methods.
The first step is to compute a nested basis of the Krylov spaces \eqref{krylovSpace}. This can be achieved using the Arnoldi algorithm (see \eg \cite{Saad2003} for a review). It simplifies for the special situation of Hermitian $H$, which we consider, and in this case is also known as Lanczos algorithm.  With the initialization
\begin{equation}
	v_1:=v(0) \, , \hspace{20pt} h_{0,1} = 0 \, , \hspace{20pt} v_0 = 0 \, ,
\end{equation}
we calculate for $j=1,\ldots m$:
\begin{subequations} \label{arnoldi}
	\begin{align}
		w_j = &H v_j \,, \\
		w_j = &w_j -  h_{j-1,j} v_{j-1}  \,, \\
		h_{j,j} =& \langle w_j, v_j\rangle  \,,\\
		\tilde{w}_j =& w_j -  h_{j,j} v_j \,,
		\label{arnoldiOrthogonalization2}\\
		h_{j,j+1} = & h_{j+1,j} =  \sqrt{\langle \tilde{w}_j, \tilde{w}_j \rangle} \,, \label{HessenbergEntry}\\
		v_{j+1} =& \tilde{w}_j/	h_{j+1,j} \,,
	\end{align}
\end{subequations}
where we use modified Gram-Schmidt orthogonalization since it is more stable numerically (see \eg \cite{Parlett1998,Higham2002,Saad2003}). As discussed in the introduction, it is important to note that no matrix-matrix multiplications $H*H$ appear but only much cheaper matrix-vector multiplications $H*v_j$. Using the Arnoldi algorithm \eqref{arnoldi}, we determine an orthogonal matrix $V_m := \left[v_1,\ldots v_m\right]$ as well as the Hermitian matrix $(H_m)_{i,j}:= h_{i,j}$. In the generic case, in which $H$ not necessarily Hermitian, (and with an appropriate generalization of \eq \eqref{HessenbergEntry}), $H_m$ is an upper Hessenberg matrix, \ie $h_{i,j} = 0$ for $i>j+1$. For Hermitian $H$, which we consider, $H_m$ is tridiagonal. Nevertheless, we shall refer to $H_m$ as the Hessenberg matrix in what follows. The matrices $H_m$ and $V_m$ fulfill
\begin{equation} \label{hessenbergEquation}
	H V_m = V_m H_m + h_{m+1,m}v_{m+1}e_m^T\,,
\end{equation}
where $e_m$ is the unit vector with entry in the $m^{th}$ component. This implies that
\begin{equation}
	H_m = V_m^T H V_m \,,
\end{equation}
\ie $H_m$ is the projection of $H$ on the subspace $\mathcal{K}_m$, in line with \eq \eqref{projection}. So far, all statements have been exact. Now we implement the approximation \eqref{approximationSymbolic}, which amounts to using $H_m$ instead of the full $H$. Thus, the approximate result of time evolving  $v(0)$ is
\begin{equation} \label{krylovApproximation}
	\tilde{v}(t) = V_m \ex^{-i H_m t}e_1 \,.
\end{equation}
This coincides with \eq \eqref{KrylovResultSymbolic} since $V_m^T v(0) = e_1$.
Here and in the following we assume that the initial state vector $v(0)$ is normalized: $||v(0)||=1$.
The key point is that unlike $H$, the Hessenberg matrix $H_m$ is small. Therefore, any standard algorithm can be used to exponentiate it without significant computational cost.

\subsection{Previous Work on Error Bounds}
Next we shall discuss bounds on the error, $\text{err}(t) := v(t)-\tilde{v}(t)$, \ie the difference of $\tilde{v}(t)$ and the result $v(t)$ of an exact time evolution (see \eq \eqref{error}). A series representation of it was derived in \cite{Saad1992}:
\begin{equation} \label{errorSeries}
	\text{err}(t) = h_{m+1,m} \sum_{k=1}^\infty \left(e_m^T \Phi_k(-i H_m t) e_1\right) \left(-i H t\right)^{k-1} v_{m+1} \;,
\end{equation}
where $\Phi_1(z) = (\ex^z - 1)/z$ and $\Phi_k(z)$ can be defined recursively (see \cite{Saad1992}). It was proposed in \cite{Saad1992} to use the first term of the series as error estimate:\footnote
{Further justifications for using \eq \eqref{errorEstimate1} as error estimate were given in \cite{Hochbruck1998}.}
\begin{equation} \label{errorEstimate1}
	||\text{err}(t)|| \approx \left|h_{m+1,m}  \left(e_m^T \Phi_1(-i H_m t) e_1\right)\right| \;.
\end{equation}
Additionally, the suggestion was made to further simply \eq \eqref{errorEstimate1} by replacing $\Phi_1(-i H_m t)$ with $\ex^{-i H_m t}$, which leads to \cite{Saad1992}:
\begin{equation} \label{errorEstimate2}
	||\text{err}(t)|| \approx \left|h_{m+1,m}  \left(e_m^T \ex^{-i H_m t} e_1\right)\right| \;.
\end{equation}
The error analysis employed in \textit{Expokit} is based on the series representation \eqref{errorSeries}.\footnote
{We remark that \textit{Expokit} employs a slightly modified version of the Krylov method \cite{Sidje1998}. In this approach, which was suggested in \cite{Saad1992}, \eq \eqref{krylovApproximation} is altered so that it also uses the vector $v_{m+1}$. In contrast, we shall stick in the present work to the original Krylov scheme as shown in \eq \eqref{krylovApproximation}. Therefore, all subsequent statement about error bounds refer to this unmodified Krylov method, but analogous results can also be obtained for the modified approach (see in particular \cite{Saad1992}).}
Following the proposals \cite{Sidje1994, Philippe1995}, both the first term (shown in \eq \eqref{errorEstimate1}) and the second term of the series \eqref{errorSeries} are used to compute an error estimate \cite{Sidje1998}. An important advantage of this approach is that it can also be applied in the general case when $H$ is not Hermitian.

A complementary approach for studying the error of the Krylov method was suggested in \cite{Lubich2008} and subsequently investigated in \cite{Botchev2013, Ye2013, Jia2015, Wang2016, Jawecki2018}. One can start from the observation that $\text{err}(t)$ fulfills a simple differential equation \cite{Celledoni1997, Druskin1998, Lubich2008, Botchev2013, Ye2013, Jia2015,  Wang2016, Jawecki2018}. In order to derive it, we first note that
\begin{equation}
	\tilde{v}'(t) = -i V_m H_m \ex^{-i H_m t}e_1 = -i H \tilde{v}(t) + i h_{m+1,m} \left( e_m^T \ex^{-i H_m t}e_1 \right) v_{m+1}\,,
\end{equation}
where we used \eqref{hessenbergEquation} in the second step.
Therefore, it follows that
\begin{equation}
	\text{err}'(t) = -iH \, \text{err}(t) -  i h_{m+1,m}\left ( e_m^T \ex^{-i H_m t}e_1 \right) v_{m+1}\,.
\end{equation}
For the initial condition $\text{err}(0)=0$, this differential equation is solved by
\begin{equation} \label{errorSolution}
	\text{err}(t) = -i h_{m+1,m} \limitint_0^t \left( e_m^T \ex^{-i H_m \tau}e_1 \right) \ex^{-i(t-\tau)H} v_{m+1} \diff \tau \,.
\end{equation}
In order to find an upper bound for the norm of $\text{err}(t)$ from \eqref{errorSolution}, one has to evaluate $||\ex^{-i(t-\tau)H}v_{m+1}||$. For a generic, \ie non-Hermitian, matrix $H$ this is very hard since it involves the exponential of a large matrix. Calculating it is as difficult as the initial problem \eqref{vOfT} we set out to solve. Thus, formula \eqref{errorSolution} makes evident the typical difficulty of rigorous error bounds: It is difficult to compute the bounding function numerically.

This problem disappears, however, when $H$ is Hermitian. Then $||\ex^{-i(t-\tau)H}v_{m+1}||=||v_{m+1}||=1$ and the bound \eq \eqref{errorSolution} can be computed explicitly. It is very interesting that the case of a Hermitian matrix $H$, which is of extraordinary relevance in physics, leads to such simplifications. We arrive at the error bound \cite{Lubich2008, Wang2016, Jawecki2018}
\begin{equation}\label{errorIntegral}
	||\text{err}(t)|| \leq \limitint_0^t \diff\tau \left| h_{m+1,m}  (e_m^T \ex^{-i H_m \tau}e_1)\right| \,.
\end{equation}
The key point of this formula is that it only depends on quantities that are known in the Krylov algorithm or easy to determine: The element $h_{m+1,m}$ is already computed in the Arnoldi procedure and $e_m^T \ex^{-i H_m \tau}e_1$ can be calculated without any significant computational cost since $H_m$ is already known and -- 
as said before -- the time needed to exponentiate the small matrix $H_m$ is negligible. 

The only remaining question is how the integral in \eq \eqref{errorIntegral} can be computed in practice. In \cite{Lubich2008} the proposal was made to approximate it by evaluating the integrand at a small number of points. For example, the right-endpoint rectangle rule represents an alternative way to arrive at the estimate \eqref{errorEstimate2} while the Simpson rule is expected to yield a more accurate approximation \cite{Lubich2008}. Additionally, it was suggested to consider the limit of a small step size, in which the integral can be performed analytically \cite{Jawecki2018}. Evidently, a major drawback of such approximation methods is that they introduce an unquantifiable uncertainty in formula \eqref{errorIntegral} so that it no longer yields a rigorous error bound.

\subsection{Improvement of Error Bound}
Whereas \textit{Expokit} relies on the series representation \eqref{errorSeries} for estimating the accuracy of the Krylov approximation, we shall in the present work use the integral \eqref{errorIntegral} for bounding the error. For computing it numerically, we will employ well-known double-exponential integration methods \cite{takahasi1974double}. In our case of a one-dimensional continuous function on a finite interval, they are known to achieve high accuracy at moderate computational costs (see \eg \cite{Davis1984, Mori2001, Bailey2005}). Dynamically adapting the number of points at which the integrand is evaluated, we can largely avoid uncertainties that would result from replacing the integral \eqref{errorIntegral} by simple approximations, such as the ones considered in \cite{Lubich2008}. In this way, inaccuracies due to the one-dimensional numerical integration become negligible as compared to other sources of error, which we shall discuss shortly.

In general, we can distinguish two reasons why the Krylov method discussed in the present work does not yield an exact result. The first one, which we have discussed so far, is due to the Krylov approximation itself, \ie the replacement of the full Hamiltonian $H$ by its projection $H_m$ on the Krylov subspace (see \eq \eqref{projection}). We can call this error ``analytic'' since it would arise even if all calculations were performed with infinite machine precision. As shown in \eq \eqref{errorIntegral}, one can provide a rigorous formula for computing a posteriori bound on this analytic error. In any numerical procedure, however, there is an inevitable second source of inaccuracy caused by a finite machine precision -- one can label the resulting error as ``numerical''. We will not bound it rigorously but only provide an estimate.

Generically, the effect of a finite machine precision is most significant when a large number of numerical operations is applied consecutively (see \eg \cite{Parlett1998,Higham2002}). Correspondingly, we expect the Arnoldi/Lanczos algorithm to be most important in the determination of the numerical error. The rounding error in the Lanczos procedure has been studied intensely and rigorous bounds were derived \cite{Kaniel1966, Paige1972, Paige1976, Paige1980}. It follows from these analyses that the order of magnitude of the numerical error can be bounded in terms of the two quantities
	\begin{equation}  \label{numericalError1}
	d\, ||H||\, \epsilon \;, \qquad \, m ||\,|H|\,||\, \epsilon \;,
\end{equation}
	where $||H||$ is the 2-norm of the matrix $H$ and $\epsilon$ is the machine precision. Moreover, $|H|$ denotes a matrix obtained from $H$ by taking the absolute value of each element. A priori the two number in \eq \eqref{numericalError1} are independent but a common bound for both of them is given by
	\begin{equation}  \label{numericalError}
	d\, ||H||_1\, \epsilon \;, 
\end{equation}
where $||H||_1$ denotes the 1-norm of $H$ and we used H\"older's inequality for a symmetric matrix. We shall use the quantity \eqref{numericalError} as an estimate for the numerical error.

In summary, we develop further existing approaches to error analysis in Krylov methods, and in particular the implementation in \textit{Expokit} \cite{Sidje1998}, in two ways. First, we use an exact formula for computing the error resulting from the Krylov approximation. It is displayed in \eq \eqref{errorIntegral} and can be evaluated in practice with only negligible additional computational effort. Apart from an insignificant inaccuracy due to a finite step size in numerical integration, formula \eq \eqref{errorIntegral} would lead to a rigorous a posteriori error bound if machine precision were infinite. Secondly, we give an estimate for roundoff errors, which result from  finite machine precision. It is shown in \eq \eqref{numericalError}. We must mention, however, that our approach is only applicable to the special case of a Hermitian matrix $H$ and therefore significantly less general than the method  employed in \textit{Expokit}.

Finally, we remark that formula \eqref{errorIntegral} does not only give a rigorous formula for computing an upper bound on the error of the Krylov approximation for a given time interval $t$. It can also be used to find the optimal step size for a given desired accuracy. As explained above, the step size denotes the time after which the Krylov procedure is restarted, \ie a new Krylov subspace is computed. The input data is the time $t$, at which we wish to evaluate $v(t)$, and an upper bound $\text{err}_{\text{max}}$ on the error. This determines an upper bound on the error rate $\text{tol}_{\text{rate}}$ as follows:
	\begin{equation} \label{errorRate}
		\text{tol}_{\text{rate}} = \frac{\text{err}_{\text{max}}}{t} \,.
	\end{equation}
	Now we can evaluate the error formula \eqref{errorIntegral} at different times. The optimal step size $	t_{\text{step}}$ is the largest time for which the resulting error rate is still below the desired bound:
	\begin{equation} \label{stepSize}
		t_{\text{step}} = \max_{\text{err}(\tilde{t})/\tilde{t} \,\leq\, \text{tol}_{\text{rate}}} {\tilde{t}} \,.
	\end{equation}
	This allows us to use a minimal number of steps while obeying a given bound $\text{err}_{\text{max}}$ on the error.

\subsection{Pseudocode}
We shall discuss more concretely our implementation of the algorithm sketched so far. It is displayed in pseudocode \ref{krylovMethod} and consists of repeating three steps: 
\begin{enumerate}
		\item We perform the Arnoldi algorithm as shown in \eq \eqref{arnoldi}. We remark that instead of $H$, we use the matrix $A=-iH$ for computing the Krylov subspace. Consequently, the Hessenberg matrix is anti-Hermitian. Otherwise, this part is completely standard. The only aspect we have not gone into so far is the special situation of a ``lucky breakdown''. Namely it can occur that the Krylov space $\mathcal{K}_m$, as defined in \eq \eqref{krylovSpace}, has a dimension that is smaller than $m$. In this case, an exact analytic result of orthonormalization, as displayed in line \eqref{arnoldiOrthogonalization2}, would yield a vanishing $h_{j+1,j}$ for some value of $j$. In the numerical implementation, this corresponds to $h_{j+1,j}$ being below some small threshold set by the machine precision. In such a case working in the Krylov subspace is no longer an approximation but yields an exact result. This means that our problem is solved and we can stop the algorithm. We shall not further discuss this special case.
	\item We find the optimal step size according to \eq \eqref{stepSize} by computing the integral in the error formula \eqref{errorIntegral} for different times $t_{\text{step}}$. We employ a double-exponential method \cite{takahasi1974double} in the form of a $\text{tanh}$-$\text{sinh}$ transformation of the integral, which is thereafter evaluated by trapezoidal quadrature. 
 In the concrete implementation, it has proven efficient to proceed as follows. In all iterations but the first one, we can use the previous step size as estimate for $t_{\text{step}}$, where we slightly reduce its value. Then we compute the resulting error according to the integral \eqref{errorIntegral} and compare it to $tol_{\text{rate}} \cdot t_{\text{step}}$. If it is larger, we half $t_{\text{step}}$ and repeat the procedure until the error is small enough. By using a slightly smaller value than the previous step size as initial estimate for $t_{\text{step}}$, we achieve that usually no halving of $t_{\text{step}}$ is required. Finally, we increase $t_{\text{step}}$ in small substeps as long as the resulting error is still small enough.
	\item Using the step size $t_{\text{step}}$ determined above, we compute $v(t_{\text{step}})$ according to \eq \eqref{krylovApproximation}. At this point, we restart the Krylov procedure, \ie we go back to step 1, where $v(t_{\text{step}})$ now is the initial vector for the next iteration of the algorithm. 
\end{enumerate}

\begin{algorithm}
	\vspace{-6\baselineskip}
	\KwIn{Hamiltonian matrix $H$,  normalized initial vector $v$, time $t$, maximal error $err_{\text{max}}$}
	\KwOut{Approximation for $ \exp(-iHt) v$} 
	
	$A := -i H$; $t_{\text{now}} :=0$; $w:=v$; $t_{\text{step}} = 0.1$; $tol_{\text{rate}}=err_{\text{max}}/t$\; 
	\While{$t_{now} < t$}
	{
		
		{//Step 1: create Krylov space}
		
		$v_1 := w$\;
		\For{$j:=1:m$}
		{
			$p:=A v_j$\;
			\If{$j\ne0$}
			{
				$p := p - M_{j-1,j} v_{j-1}$\;	 // $M$ is different name for Hessenberg matrix $H_m$	\\
			}
			$M_{jj} := v_j  p$; 
			$p:=p-M_{jj} v_j$\;
			$no:= ||p||_2 $\;
			\If{$no < tol_{\text{rate}}$} {// lucky breakdown}
			\uIf{$j \neq m$}
			{
				$M_{j,j+1}:=-no$; $M_{j+1,j}:=no$\;
				$v_{j+1}:=p/no$\;
			}
			\Else
			{
				$h:=no$\;
			}
		}
		
		//Step 2: find optimal step size
		
	$f(t) := h\, |e_m^T \cdot \exp(M \cdot t) \cdot e_1|$;	$t_{\text{step}} := 0.97\, t_{\text{step}}$ \;
	$err_{\text{step}} :=\text{integrate}\  f(t) \ \text{from} \ 0 \ \text{to}\ t_{\text{step}}$\;
			\While{$err_{\text{step}} > tol_{\text{rate}} \cdot t_{\text{step}}$}
		{
			$t_{\text{step}}= t_{\text{step}}/2$\;
			$err_{\text{step}} :=  \text{integrate}\  f(t) \ \text{from} \ 0 \ \text{to}\ t_{\text{step}}$\;
		}
	$s = t_{\text{step}} / \text{N\_SUBSTEPS}$; $n_s=0$; $\Delta_{\text{err}}=0$\;
	\While{$err_{\text{step}} + \Delta_{\text{err}} < tol_{\text{rate}} \cdot (t_{\text{step}}+n_s*s) $} 
	{
		 $err_{\text{step}} =err_{\text{step}} +  \Delta_{\text{err}}$; $n_s = n_s+1$\;
		 $\Delta_{\text{err}} =   \text{integrate}\  f(t) \ \text{from} \ t_{\text{step}}+n_s*s \ \text{to}\ t_{\text{step}}+(n_s+1)*s$\;
	}
$t_{\text{step}} = t_{\text{step}} + (n_s-1)\cdot s$\;
		$\omega:= \exp(M \cdot t_{\text{step}} \cdot e_1)$\;
	
		//Step 3: perform step
		
		$w = V \cdot \omega$;	$t_{\text{now}} := t_{\text{now}} + t_{\text{step}}$\;
	}
	\caption{Krylov-method}
	\label{krylovMethod}
\end{algorithm}

For clarity of presentation, a number of additional functionalities and special cases are not shown in the pseudocode \ref{krylovMethod}:
\begin{itemize}
	\item It is important to point out that the fraction of runtime required for step 2 is generically small since computations are only performed in the small Krylov subspace of dimension $m$. Nevertheless, one can gain a slight improvement in performance by diagonalizing the Hessenberg matrix $H_m$ at the beginning of step 2 (see \eg \cite{Lubich2008}). If this is done once, no matrix exponential is needed any more to compute the integrand of \eqref{errorIntegral} at different values of its argument. We use this approach in our implementation.
	\item In the first iteration, we have no previous step size which we could use as estimate for the next one.
	 Therefore, we slightly modify step 2 as follows. With an arbitrary initial guess for $t_{\text{step}}$, we compute the error. If it is small enough, we repeatedly double $t_{\text{step}}$ until the next doubling would make the error too large. Then we proceed with the displayed procedure, \ie we first keep halving $t_{\text{step}}$ if the error is too big and subsequently increase $t_{\text{step}}$ in small substeps.
	\item At the end of step 2, we compare the error computed according to \eq \eqref{errorIntegral} with the estimate \eqref{numericalError} of the numerical error. If the latter is bigger, the program triggers a warning to alert the user that the analytic error bound may be spoiled by finite machine precision.
	\item The above algorithm can trivially be extended not only to compute the final vector but also to sample its values at intermediate points of time: In step 3, one uses the known matrices $H_m$ and $V_m$ to compute the vector at any $t_s$ with $t_{\text{now}}\leq t_s \leq t_{\text{now}}+t_{\text{step}}$.
\end{itemize}

\section{Creation of Hamiltonian Matrix}
\label{sec:numberOperators}
So far, we have shown how time evolution can be computed once the Hamiltonian matrix $H$ is known. A problem that one encounters in practical applications is that the matrix $H$ is usually not given. Instead, the physical system and the Hamiltonian are defined in more abstract terms. In order to demonstrate how the matrix $H$ can be derived in such a situation, we shall focus on the widely-used description of physical systems in terms of number operators.  We shall brief\/ly review it and then discuss how H can be derived.

Every physical system can be described as collection of interacting oscillator modes. We take their number to be $L$ and assume that $L$ is finite.\footnote
{Otherwise the Hilbert space has infinite dimension and our numerical approach is not applicable.}
Each mode $l$ possesses creation and annihilation operators $\hat{a}_l^\dagger$, $\hat{a}_l$, where $l =1,\ldots,L$. In the case of bosons, they fulfill the algebra
\begin{equation} 
	[\hat{a}_k,\hat{a}_l^{\dagger}] = \delta_{kl}\,,\qquad
	[\hat{a}_k,\hat{a}_l]  =   [\hat{a}_k^{\dagger},\hat{a}_l^{\dagger}] =0\,,   
	\label{algebra} 
\end{equation}
and lead to the number operators
\begin{equation}
	\hat{n}_l = \hat{a}_l^\dagger \hat{a}_l \,. 
\end{equation}
Now we can use the creation operators to form a basis of states:
\begin{equation} \label{basisState}
	\ket{n_1,\ldots, n_L} := \left(\hat{a}_1^\dagger\right)^{n_1} \ldots \left(\hat{a}_L^\dagger\right)^{n_L} \ket{0}\,,
\end{equation}
where $\ket{0}$ is the vacuum and the numbers $(n_1,\ldots, n_L)$ characterize the occupation number of the different modes:
\begin{equation}
	\hat{n}_l \ket{n_1,\ldots, n_L} = n_l \ket{n_1,\ldots, n_L}\,.
\end{equation}
In the non-relativistic limit, the total occupation number,
\begin{equation} \label{totalOccupationNumber}
	N = \sum_{l=1}^L n_l\,,
\end{equation}
is conserved. In this case, not all tuples $(n_1,\ldots, n_L)$ are admissible and the set of states \eqref{basisState} is finite. 
Finally, the Hamiltonian can be expressed as an operator in terms of the creation and annihilation operators. For example, the free Hamiltonian for a system of $L$ modes is given by
\begin{equation}
	\hat{H}_0 = \sum_{l=1}^L \varepsilon_l \hat{a}_l^\dagger \hat{a}_l \,,
\end{equation}
where $\varepsilon_l$ are arbitrary real numbers.

In order to be able to apply \textit{TimeEvolver}, one now needs to derive the Hamiltonian matrix $H$ from an Hamiltonian operator $\hat{H}$, which is expressed in terms of creation and annihilation operators. As a first step, we order the basis elements, \ie we introduce an arbitrary but fixed bijective function $\text{ind}$, which maps the admissible tuples $(n_1,\ldots, n_L)$ on an index set $I=\{1,\ldots, d\}$. For each basis element $\ket{n_1,\ldots, n_L}$, we now apply the Hamiltonian operator $\hat{H}$, which in general leads to a superposition of basis elements:
\begin{equation}
	\hat{H} \ket{n_1,\ldots, n_L} = \sum_k c^{(k)} \ket{n^{(k)}_1,\ldots, n^{(k)}_L} \,.
\end{equation}
Here $c^{(k)}$ are a complex numbers.
Then we define for each non-zero $c^{(k)}$:
\begin{equation}
	H_{i,j} = c^{(k)} \,,
\end{equation}
where
\begin{equation} \label{findIndices}
	i = \text{ind}^{-1}\left(\ket{n_1,\ldots, n_L}\right) \,, \qquad j = \text{ind}^{-1}\left(\ket{n^{(k)}_1,\ldots, n^{(k)}_L}\right) \,.
\end{equation}
At last, the $H_{i,j}$ define the elements of the Hamiltonian matrix. 

Finally, we remark that the above procedure can be straightforwardly generalized to Hamiltonians that are defined in terms of other operators than creation and annihilation operators. To this end, one only needs to adapt the commutation relations \eqref{algebra} and the mapping \eqref{basisState} of basis states and operators. Finally, superselection rules, such as number conservation in our case (see \eq \eqref{totalOccupationNumber}), need to be taken into account in the enumeration of basis states.

\section{Description of Program}
\label{sec:implementation}
We have implemented our method in C++. Our program uses the following libraries. 
The core module relies on the Intel\textsuperscript{\circledR} Math Kernel Library (MKL)\footnote{\url{https://software.intel.com/content/www/us/en/develop/tools/oneapi/components/onemkl.html}} for linear algebra operations.\footnote{We choose MKL because it provides LAPACK and sparse BLAS  routines in one package as well as for performance reasons.} Moreover, Boost\footnote{\url{https://www.boost.org/}} is used for numerical integration. These two libraries are the only necessary prerequisites for our program.
Outside of this core functionality, we employ Hierarchical Data Format version 5 (HDF5\textsuperscript{\circledR})\footnote{\url{https://www.hdfgroup.org/solutions/hdf5}} for convenient output of data. Finally, CMake\footnote{\url{https://cmake.org/}} is used to build our program.

Our implementation of the Krylov subspace method is contained in the class \texttt{krylovTimeEvolver} in the folder ``core''. Its constructor requires all arguments that are needed for time evolution, in particular the Hermitian Hamiltonian matrix and the complex vector of the initial state, where the corresponding data types are defined in \texttt{matrixDataTypes}. Algorithm \ref{arnoldi} is implemented in the method \texttt{timeEvolve}, where Step 1, \ie the Arnoldi algorithm, is encapsulated in \texttt{arnoldiAlgorithm} and Step 2, \ie the determination of the maximal step size compatible with the desired error bound, is performed in \texttt{findMaximalStepSize}. 

For evaluating the error integral \eqref{errorIntegral} in \texttt{findMaximalStepSize} numerically, we rely on an adaptive implementation of a double-exponential method provided by Boost. It increases the number of points, at which the integrand is sampled, until an estimate of the relative error in evaluating the integral, which is incorporated in the library, falls below $10^{-3}$. If this cannot be achieved with $15$ refinements, the program triggers a warning. In certain cases, this sophisticated integration procedure can have a negative impact on runtime. Therefore, we have also implemented a second integration method based on a non-adaptive Gau\ss-scheme (see \eg \cite{Davis1984}.), which the user can choose to use. This simpler and faster approach could be particularly useful for explorative investigations. If interesting effects are found, one can establish the validity of results with the more accurate double-exponential method. We remark, however, that the effect of integration on runtime becomes negligible for large Hilbert spaces ($d \gtrsim 10^5$) so that in this case there is no disadvantage in always using the more precise approach.

All core functionalities of \textit{TimeEvolver} are contained in the \texttt{krylovTimeEvolver}-class. If users already have at their disposal a Hamiltonian matrix, no parts of our program other than the class \texttt{krylovTimeEvolver} are relevant for them.
However, for the convenience of the user who does not yet have a Hamiltonian matrix, we have also added the folder ``helper", which contains routines to form it. The class \texttt{basis} provides a representation of basis vectors in terms of number eigenstates, as described in section \ref{sec:numberOperators}. In particular, its creator forms all basis elements for a given number $K$ of modes and a given total occupation $N$. Moreover, it supports the case in which some of the modes have a given fixed maximal occupation number.

The class \texttt{hamiltonian} contains a representation of a Hamiltonian in terms of creation and annihilation operators and provides the important routine \texttt{createMatrix}, which creates a Hamiltonian matrix according to the approach described in section \ref{sec:numberOperators}. Since the function $\text{ind}$ is inverted frequently (see \eq \eqref{findIndices}), we represent it by a hashtable, which can perform this operation in constant time. 
We remark that the routine \texttt{createMatrix} can be easily adapted to other choices of basis. This makes it possible to apply it to Hamiltonians that are not defined in terms of creation and annihilation operators.

Finally, we have included the folder ``example", which demonstrates the usage of \textit{TimeEvolver} for a particular class of Hamiltonians. These Hamiltonians were investigated in the study \cite{Dvali2020}, which relies on the \textit{TimeEvolver} for numerical time evolution.\footnote
{More precisely, the prospect of using the \textit{TimeEvolver} in the study \cite{Dvali2020} inspired us to create it.}
The particular structure of the Hamiltonians is reflected in the class \texttt{exampleHamiltonian}. The routine \texttt{main} contained in the class of the same name shows how all parts can be integrated and the \textit{TimeEvolver} is used. Results of this routine are shown in section \ref{ssec:exemplaryUsage}.

\section{Application of Program}
\label{sec:performance}

\subsection{Exemplary System}
\label{ssec:exemplaryUsage}

We will demonstrate the usage of \textit{TimeEvolver} on the specific Hamiltonian used in \cite{Dvali2020}, which we shall brief\/ly describe. The system consists of two subsystems of quantum oscillators. The first set is comprised of the two modes $\hat{a}_0$ and $\hat{b}_0$. We label their occupation numbers as $n_0$ and $m_0$. The second subsystem includes the modes $\hat{a}_k$ and $\hat{a}'_{k'}$, where $k= 1,\dots, K$ and $k'= 1,\dots, K'$. These oscillators are assumed to be qubits, \ie their occupation numbers $n_k$ and $n'_{k'}$ are restricted to the values $0$ and $1$. In total the system features $L = 2 + K + K'$ modes. We can denote a number eigenstate as
	\begin{equation} \label{genericStateExample}
	\ket{n_0,m_0,n_1, \ldots, n_K, n'_1, \ldots, n'_{K'}} \,.
	\end{equation}
As will become evident, the system is constructed in such a way that the occupation numbers in the two sets of oscillators are conserved independently. Thus, the two numbers
\begin{equation} \label{subsetOccupations}
	N_0 = n_0 + m_0 \,, \qquad N_m = \sum_{k =1}^K n_k + \sum_{k'=1}^{K'} n'_{k'}
\end{equation}
do not change in the course of time evolution.

Apart from this property, the particular structure of the system is not important for the present study. For completeness, we shall nevertheless discuss it in more detail and outline its physical motivation. The concrete Hamiltonian is given by
\begin{align}
	\hat{H} =  C_0& \left(  \hat{a}_0^\dagger \hat{b}_0 + \hat{b}_0^\dagger \hat{a}_0\right) 
	+  \epsilon_m\left(1-\frac{\hat{n}_0}{N_c}\right) \sum_{k =1}^K \hat{n}_k 
	\nonumber\\ + 
	\epsilon_m &\left(  1-\frac{\hat{n}_0}{N_c-\Delta N_c}\right) \sum_{k'=1}^{K'}  \hat{n}'_{k'} 
	\nonumber\\ + 
	C_{\text{m}} &\Bigg\{ \sum_{k=1}^{K}\sum_{k'=1}^{K'} f_1(k,k')\left( \hat{a}_k^\dagger \hat{a}'_{k'} +  \text{h.c.}\right) 
	\nonumber\\
	&+ \sum_{k=1}^{K}\sum_{\substack{l=1\\l> k}}^{K} f_2(k,l)\left( \hat{a}_k^\dagger \hat{a}_l +  \text{h.c.}\right) 
	\nonumber\\ 
	&+\sum_{k'=1}^{K'}\sum_{\substack{l'=1\\l'> k'}}^{K'}  f_3(k',l')\left( \hat{a}_{k'}^{'\dagger} \hat{a}'_{l'} +  \text{h.c.}\right)\Bigg\} \,,
	\label{fullHamiltonianSimple}
\end{align}
with 
\begin{equation}
	f_i(k,l) = 
	\left\lbrace \begin{array}{rcl}
		F_i(k,l) -1 & \mbox{for} & F_i < 0.5
		\\ 
		F_i(k,l) & \mbox{for} & F_i\geq 0.5 
	\end{array}\right. \,,
\end{equation}
where $F_i(k,l) = \left( \sqrt{2}(k+\Delta k_i)^3 + \sqrt{7}(l+\Delta l_i)^5 \right) \mod 1$. Moreover, we set $\Delta k_1 = \Delta k_2 = 1$, $\Delta k_3 = K+1$, $\Delta l_1 = \Delta l_3 = K+1$ as well as $\Delta l_2 = 1$. The Hamiltonian $\hat{H}$ depends on the real constants $\epsilon_m$, $C_0$, $N_c$, $\Delta N_c$, $C_m$, $K$ as well as $K'$, which fulfill $N_c >0$, $0 < \Delta N_c < N_c$, $K>0$ and $K'>0$.
As initial state, we use
\begin{equation} \label{initialStateSimple}
	\ket{\text{in}}=\ket{N_0,0,\underbrace{1,\ldots, 1}_{N_m}, 0, \ldots, 0} \,,
\end{equation}
where we introduced $N_0$ and $N_m$ as additional parameters. They denote the number of particles in the mode $\hat{a}_0$ and the number of qubits that are set to $1$, respectively. In accordance with \eq \eqref{subsetOccupations}, they moreover determine the number of particles in each of the two subsectors of the system. Unless specified otherwise, we use the following choice of parameters:
\begin{align} \label{systemParameters}
	\epsilon_m = \sqrt{20} \,, \quad N_0 = N_c=20\,, \quad  \Delta N_c = 12  \,, \quad K=K'=4 \,,\\ \nonumber C_0 = 1\,, \quad C_m = 1\,,\quad N_m=2 \,.
\end{align}

Finally, we shall brief\/ly explain why it is interesting to study the system \eqref{fullHamiltonianSimple}, thereby elucidating the origin of the concrete Hamiltonian. If the reader is only interested in the application of \textit{TimeEvolver}, they can skip this part. The original motivation comes from black hole physics. The corresponding line of research was initiated in \cite{Dvali:2012en} and a complete list of references can be found in \cite{Dvali2020}. A generic problem in black hole physics is that exact calculations beyond any semi-classical limit are very hard to perform. Therefore, it is interesting to look for analogue systems which share important properties with black holes, but which are easier to study. Then one can try to draw conclusions about a black hole by solving the analogue system. An important property of a black hole is its Bekenstein-Hawking entropy \cite{Bekenstein:1973ur}, because of which it has a high capacity of information storage. This motivates the study of generic quantum systems with an enhanced memory capacity.

Thus, we start from the question how a generic quantum system can achieve a high capacity of information storage. We can quantify the latter by a microstate entropy $K$, which is defined as the logarithm of the number of available states. Two ingredients are sufficient. First, one needs $K$ modes. The number of states corresponding to different occupation numbers of them (as defined in \eq \eqref{basisState}) scales exponentially with $K$. For example, if occupation numbers are restricted to $0$ and $1$, then $K$ qubits leads to $2^K$ microstates and hence an entropy on the order of $K$. Secondly, all microstates must be nearly degenerate in energy. If this is not the case, they cannot be counted as microstates that belong to one and the same macrostate. 

Both of these requirements are fulfilled by the simple system \cite{Dvali:2017ktv}
\begin{equation}
\hat{H}_{\text{prototype}} =   \epsilon_m\left(1-\frac{\hat{n}_0}{N_c}\right) \sum_{k=1}^K \hat{n}_k \,.
\end{equation}
It corresponds to a part of the Hamiltonian \eqref{fullHamiltonianSimple}, and all definitions are as there. The qubits $\hat{a}_1$, \ldots, $\hat{a}_K$ are responsible for information storage, \ie generating the microstate entropy $K$ -- we can call them \textit{memory modes}. If the mode $\hat{a}_0$ were absent, however, the states corresponding to different occupation number of the memory modes would not be degenerate in energy. Indeed, there would be a large difference of $K \epsilon_m$ between the energies of the empty and of the full state. This changes once $\hat{a}_0$ is populated. In a highly occupied state with $\braket{n_0}=n_0$, we can use the Bogoliubov approximation, $\hat{n}_0 \approx n_0$, to derive the effective energy gap of the $\hat{a}_k$-modes:
\begin{equation} \label{effectiveGap}
\epsilon_{\text{eff}} \approx \left(1-\frac{n_0}{N_c}\right) \epsilon_m \,.
\end{equation}
We see that it is lowered, and for a critical occupation $n_0 = N_c$ all memory modes become gapless. Then the microstates corresponding to different occupation numbers of them become degenerate in energy and an entropy on the order of $K$ is achieved. We call this phenomenon, in which a high occupation number of one mode decreases the energy gap of others, \textit{assisted gaplessness} \cite{Dvali:2018tqi}. This is the reason why we choose $N_0=N_c$ in the initial state \eqref{initialStateSimple}. 

These observations lead to a question about how the memory modes backreact on the time evolution of $\hat{a}_0$. In order to study it, we couple $\hat{a}_0$ to another mode $\hat{b}_0$, which results in the first line of the Hamiltonian \eqref{fullHamiltonianSimple}. If the memory modes $\hat{a}_k$ are empty, the occupation numbers $n_0$ and $m_0$ perform oscillations. Once the $\hat{a}_k$ are occupied, however, the oscillations get suppressed and $n_0$ becomes tied to the initial state $n_0 = N_c$. This effect of \textit{memory burden} \cite{Dvali:2018xpy} arises because any deviation of $n_0$ from $N_c$ would destroy the gaplessness of the memory modes and hence be very costly in energy, as is evident from \eq \eqref{effectiveGap}. 

In \cite{Dvali2020}, we studied the question if memory burden can be overcome. To this end, we introduced a second set of memory modes $\hat{a}'_{k'}$, which becomes gapless at a different occupation number $n_0 = N_c - \Delta N_c$. The corresponding operators are displayed in the second line of the Hamiltonian \eqref{fullHamiltonianSimple}. Moreover, we allowed transitions between and among the two sets of memory modes, which explains why we have the remaining lines of \eq \eqref{fullHamiltonianSimple}. In the presence of a second set of memory modes, it becomes energetically possible to transit from the initial state \eqref{initialStateSimple}, in which $n_0 = N_c$ and only the $\hat{a}_k$ are occupied, to a final state, in which $n_0 = N_c - \Delta N_c$ and only $\hat{a}'_{k'}$ are occupied. In \cite{Dvali2020}, we used \textit{TimeEvolver} to study if such transitions actually occur dynamically, and our conclusion was that they are generically heavily suppressed. Applied to black holes, this finding has interesting implications for Hawking radiation \cite{Hawking:1974sw}. Namely, it indicates that evaporation of a black hole slows down significantly at the latest after it has lost half of its mass. 

\subsection{Analysis of Performance}
Now we shall use \textit{TimeEvolver} to compute the time evolution of the initial state \eqref{initialStateSimple} with the Hamiltonian \eqref{fullHamiltonianSimple}. The corresponding code is included in the folder ``example''. The Hamiltonian is represented by the class \texttt{exampleHamiltonian} and the method \texttt{main} shows how \textit{TimeEvolver} is applied. For the numerical precision on the norm we set $\text{err}_{\text{max}} = 10^{-8}$. As exemplary observables we use the expectation values of the number operators of all quantum modes.

\begin{figure}
	\centering 
	\begin{subfigure}{0.4\textwidth}
		\includegraphics[width=\textwidth]{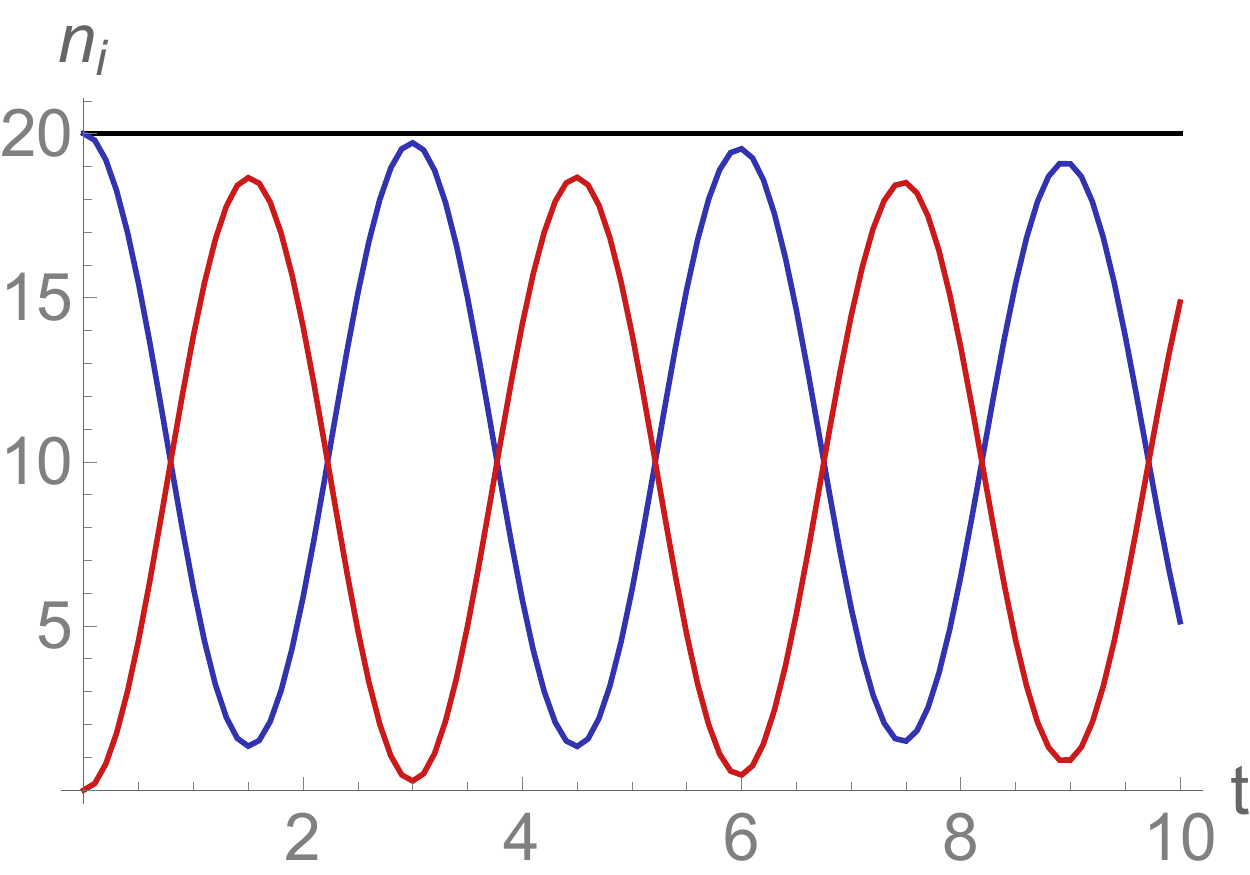}
		\caption{Expectation value of the number operators in the first sector, which consists of $\hat{a}_0$ and $\hat{b}_0$}
		\label{sfig:example1a}
	\end{subfigure}
	\hspace{0.05\textwidth}
	\begin{subfigure}{0.4\textwidth}
		\includegraphics[width=\textwidth]{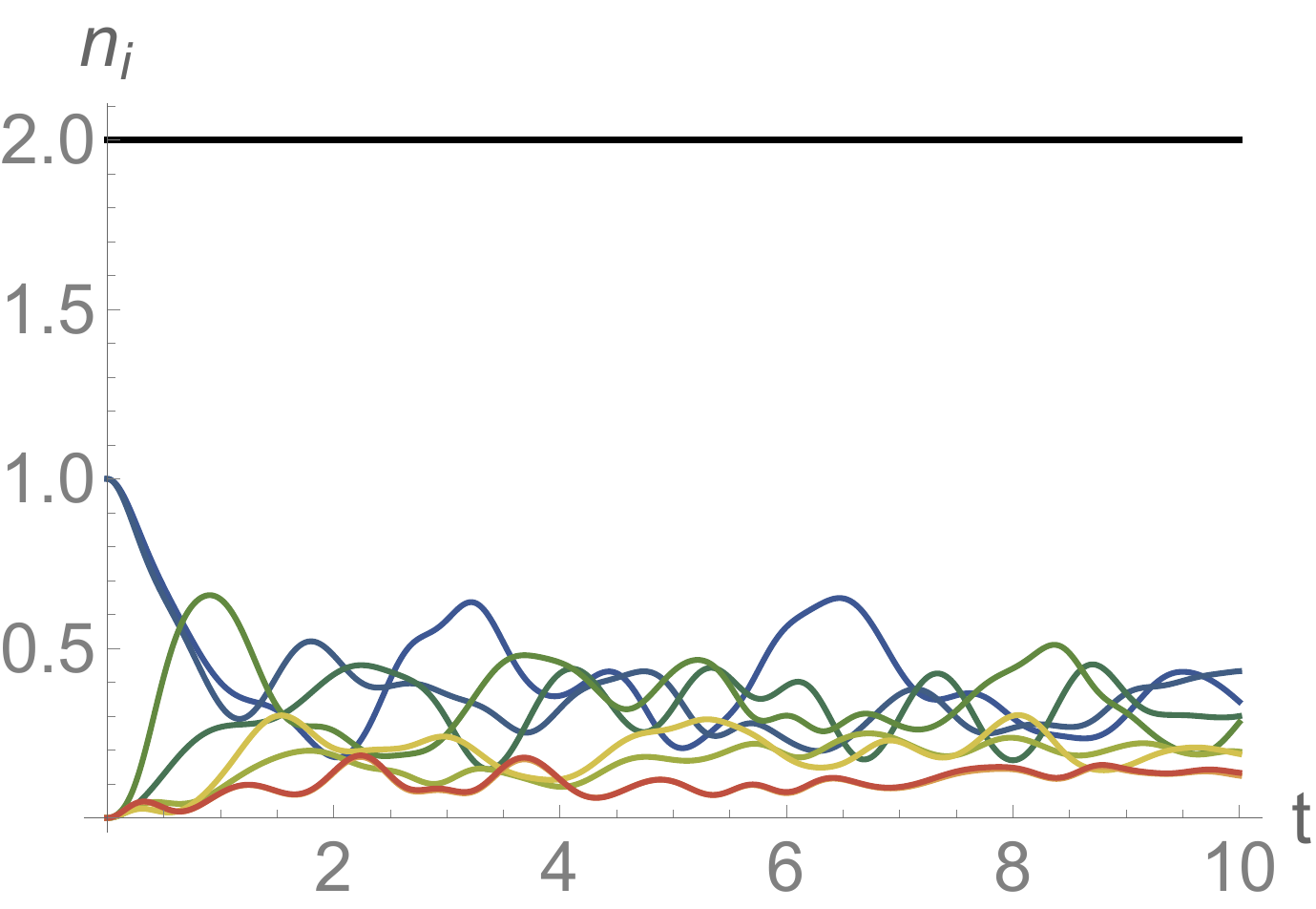}
		\caption{Expectation value of the number operators in the second sector, which consists of $\hat{a}_1,\ldots,\hat{a}_K,\hat{a}'_{1},\ldots,\hat{a}'_{K'}$}
		\label{sfig:example1b}
	\end{subfigure}
	\caption{Time evolution of the initial state \eqref{initialStateSimple}. The horizontal black line plot illustrates the total occupation numbers $N_0$ and $N_m$ in the two subsector. They are conserved separately.}
	\label{fig:example1}
\end{figure}

\begin{figure}
	\centering 
	\begin{subfigure}{0.4\textwidth}
		\includegraphics[width=\textwidth]{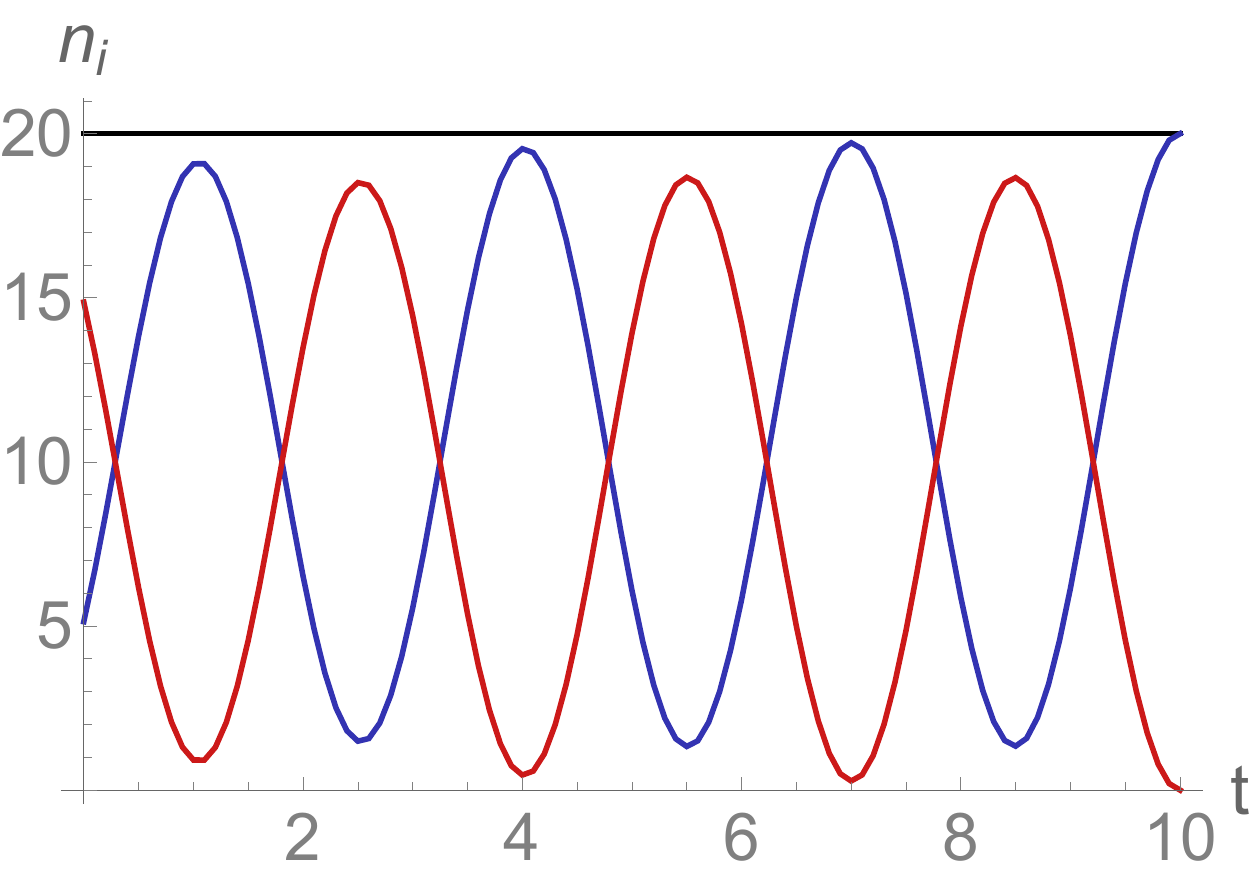}
		\caption{Expectation value of the number operators in the first sector, which consists of $\hat{a}_0$ and $\hat{b}_0$}
		\label{sfig:example1aR}
	\end{subfigure}
	\hspace{0.05\textwidth}
	\begin{subfigure}{0.4\textwidth}
		\includegraphics[width=\textwidth]{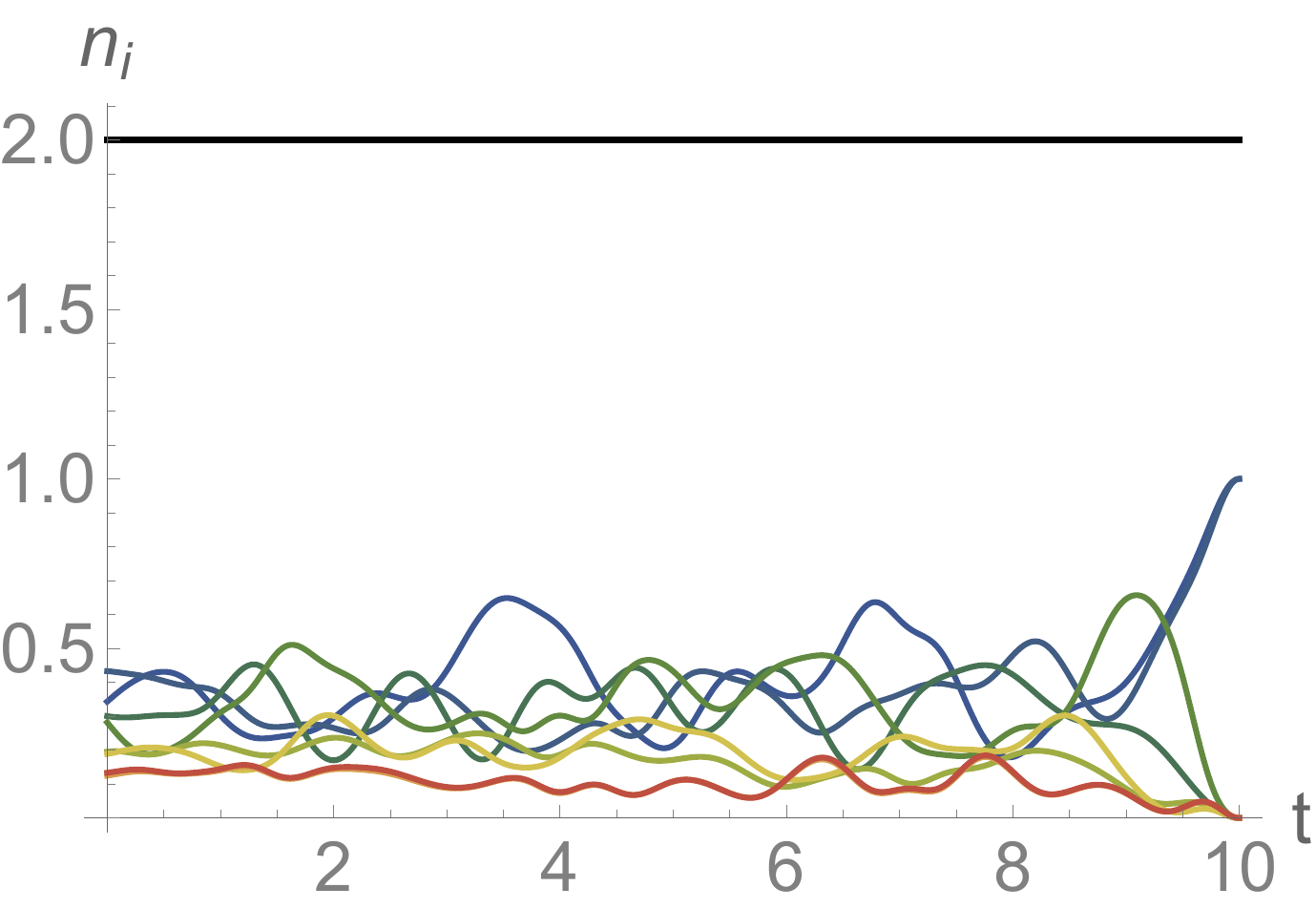}
		\caption{Expectation value of the number operators in the second sector, which consists of $\hat{a}_1,\ldots,\hat{a}_K,\hat{a}'_{1},\ldots,\hat{a}'_{K'}$}
		\label{sfig:example1bR}
	\end{subfigure}
	\caption{Time-reversed evolution of the final state shown in \fig \ref{fig:example1}.}
	\label{fig:example1R}
\end{figure}

For the choice \eqref{systemParameters} of parameters, the occupation numbers as functions of time are displayed in \fig \ref{fig:example1}. Moreover, we evaluate the sums $N_0$ and $N_m$ of occupation numbers in the two sectors of the system, as defined in \eq \eqref{subsetOccupations}. That $N_0$ and $N_m$ are indeed constant represents a first consistency check.
As a second consistency check we compute a forward-backward example. To this end, we first calculate time evolution of the state \eqref{initialStateSimple} up to $t=10$, as displayed in \fig \ref{fig:example1}, and then use the result as initial state for another time evolution, in which we replace $\hat{H} \rightarrow - \hat{H}$. \fig \ref{fig:example1R} shows the result of the second computation. In accordance with quantum mechanical unitarity, we see that \fig \ref{fig:example1R} is the mirror image of \fig \ref{fig:example1}. Moreover, we arrive again at the initial state with an error of $9.82 \times 10^{-9}$ which is within the requested error bound of $2.0 \times 10^{-8}$. 

Next, we shall analyze the runtime. All following benchmark results in this section were obtained on an Intel\textsuperscript{\textregistered} Core\texttrademark  i9-9900K Processor with 22GB RAM. First we compare the simulation wall time of the \textit{TimeEvolver} with the original \textit{Expokit} implementation (in its MATLAB version).\footnote{We used MATLAB R2021b for the Benchmark.} As a prototype system we choose \eqref{fullHamiltonianSimple} with the model parameters set to $K=K'=10$, $N_m = 5$ and $N_0=N_c=100$ and the remaining ones given by \eqref{systemParameters}. This corresponds to a Hamiltonian matrix of dimension $1,565,904\times1,565,904$. We evolve the initial state \eqref{initialStateSimple} up to $t=10$, choose for both implementations a Krylov space dimension of $m=40$ and request an upper error bound of $\text{err}_{\text{max}}=10^{-7}$. Restricted to one thread we measured the following mean simulation times averaged over five runs:
	\begin{equation}
		t_{\text{TimeEvolver}} = 353\, \text{s} \,, \hspace{20pt}  	t_{\text{Expokit}} = 1537\, \text{s} \,.
	\end{equation}
The 2-norm difference between the final states is $||v(10)_{\text{TimeEvolver}} - v(10)_{\text{Expokit}} || = 5.7 \times 10^{-8}$, and thus consistent with the error bounds. However, we remark that the estimate for the numerical error based on \eq \eqref{numericalError} yields a larger value, $3.2 \cdot 10^{-6}$, and so a warning is triggered in \textit{TimeEvolver}. This warning also appears in some of the examples discussed subsequently. Finally, we note that using the fast integration based on Gau\ss-quadrature does not lead to a measurable improvement in runtime due to a large Hilbert space dimension.

\begin{figure}
	\centering
	\includegraphics[width=0.6\textwidth]{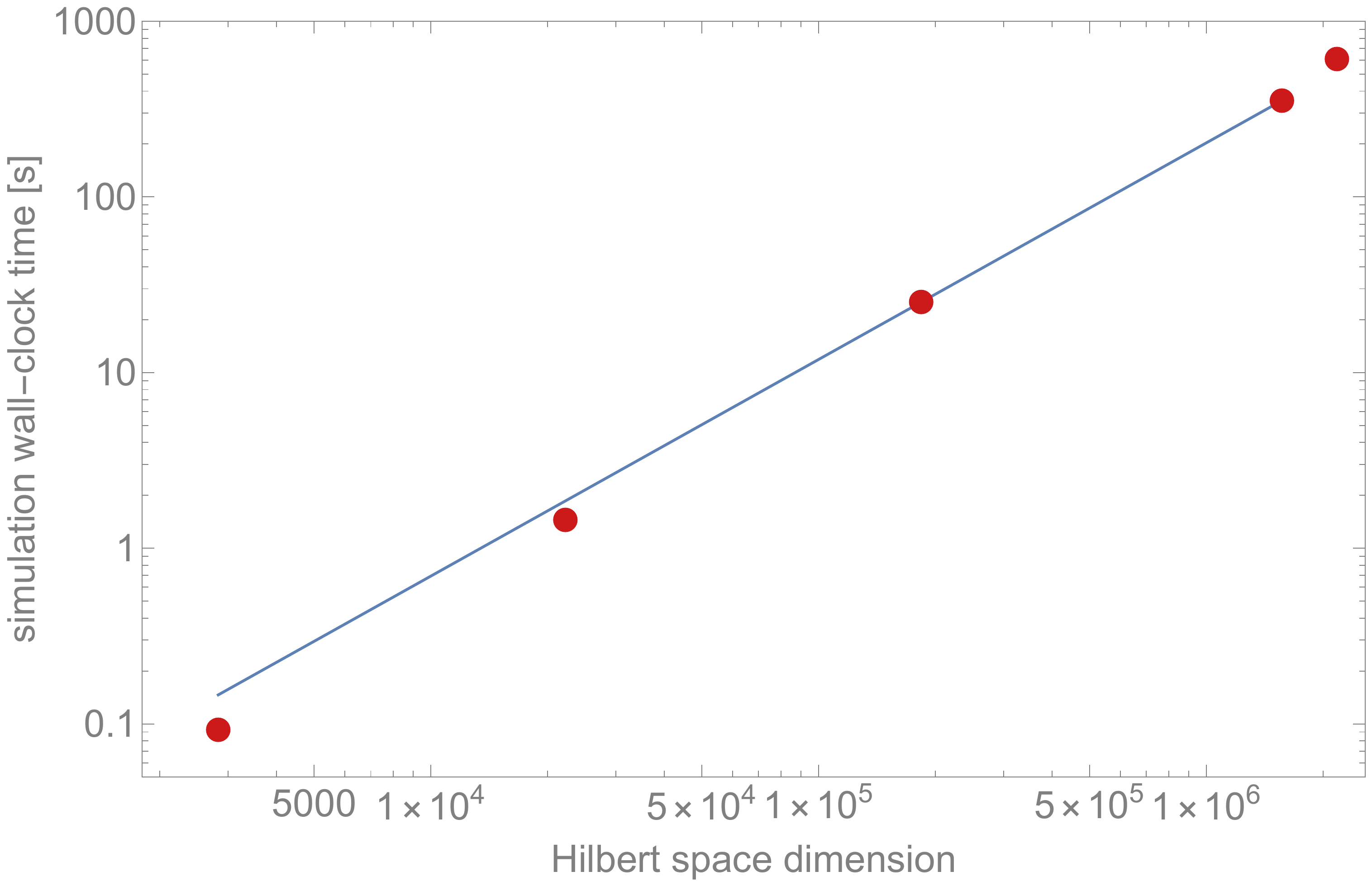}
	\caption{Simulation time as a function of Hilbert space dimension. The data points correspond to the parameter choices $N_0=N_c=100$, $K=K'=4,6,8,10$, $N_m=K/2$ and $t_{max}=10$, with the remaining ones set by \eqref{systemParameters}. The last point has been obtained by setting $K=K'=10$ and $N_0=N_c=139$ to go to the memory limit of our specific machine. The blue straight line represents a fit to the data, as described in the text. The last point has not been included in the fit.}
	\label{fig:hilbert}
\end{figure}

 Our next goal is to study how runtime depends on the size of the system. Again using the Hamiltonian \eqref{fullHamiltonianSimple} as a prototype model, we shall vary the number $K$ of modes in the memory sector. We set $K'=K$ as well as $N_m = K/2$ and $N_0=N_c=100$.
Note that the amount of basis elements depends exponentially on the number of modes. The simulation time as a function of the Hilbert space dimension is illustrated in \fig \ref{fig:hilbert}. The last data point corresponds to a Hilbert space dimension of $2,170,560$ which was the largest one fitting in the memory of the benchmark machine. We note that the limitation on the Hilbert space dimension comes from the creation of the Hamiltonian matrix and not from the actual time evolution. By making matrix creation more memory efficient, one could slightly increase the maximal dimension of Hilbert space.
	Finally, we note that for our specific prototype system the relation of Hilbert space dimension $d$ and simulation time $t$ is polynomial. Performing a fit of the form $t= a \cdot d^b$ to the data, we obtain $a=8.1 \times 10^{-6}$ and $b=1.23$, where we excluded the last point due to overhead effects of being at memory limit. The result of the fit is shown in \fig \ref{fig:hilbert} as blue straight line.

\begin{figure}
	\centering 
	\begin{subfigure}{0.4\textwidth}
		\includegraphics[width=\textwidth]{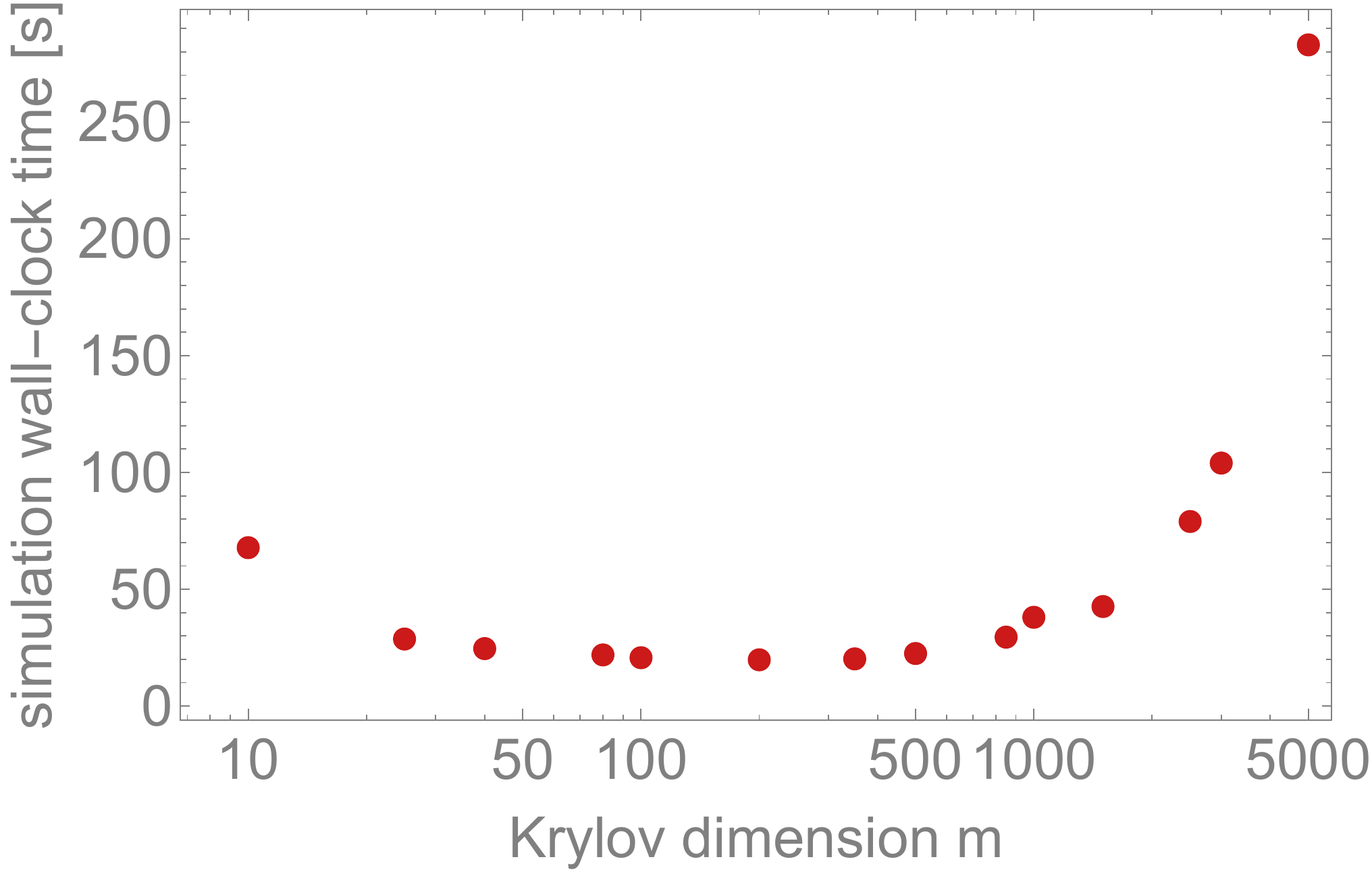}
		\caption{Simulation time scaling with Krylov dimension $m$.} 
		\label{sfig:krylov}
	\end{subfigure}
	\hspace{0.05\textwidth}
	\begin{subfigure}{0.4\textwidth}
		\includegraphics[width=\textwidth]{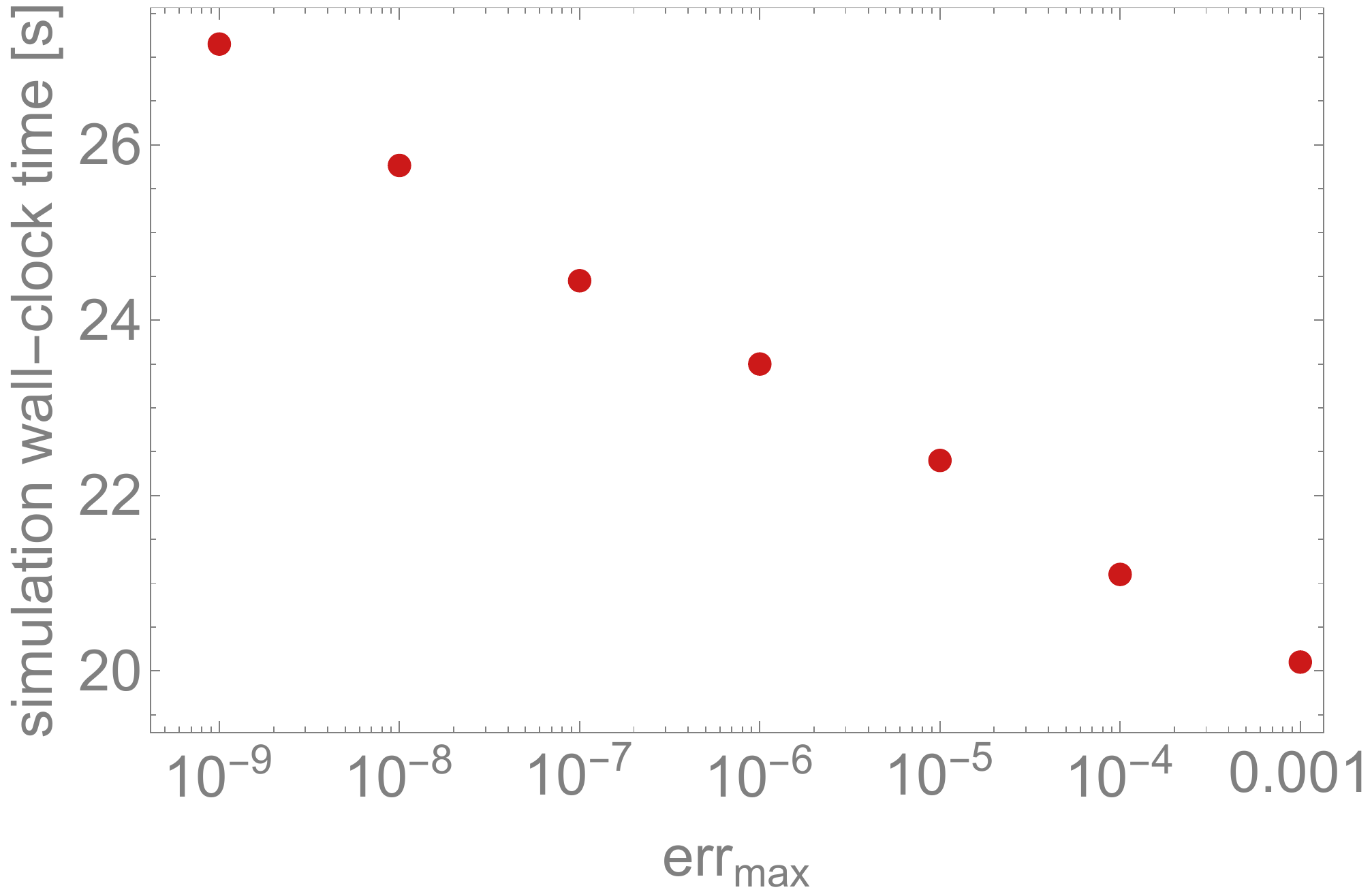}
		\caption{Simulation time scaling with requested bound $\text{err}_{\text{max}}$ on the norm error.}
		\label{sfig:norm}
	\end{subfigure}
	\caption{Wall clock time scaling. The parameter of the model \eqref{fullHamiltonianSimple} were chosen to be $K=K'=8$, $N_0=N_c=100$, $N_m=4$ and $t_{max}=10$ with the remaining ones set by \eqref{systemParameters}}
	\label{fig:performance}
\end{figure}

To conclude, we study the dependence of the simulation time on meta parameters of the algorithm itself, namely the Krylov space dimension $m$ and the requested tolerance on the norm error $\text{err}_{\text{max}}$. For this study, we set $K=K'=8$, $N_0=N_c=100$, $N_m=4$ and $t_{max}=10$ with the remaining ones set by \eqref{systemParameters}. The results are displayed in \fig \ref{fig:performance}. It is evident from \fig \ref{sfig:krylov} that the runtime only depends mildly on the dimension of Krylov space. The range of acceptable values of $m$ is usually rather large and penalty on the simulation time is in most cases not drastic as long as it is not unreasonably small or large. 
The wall-clock simulation time as a function of $\text{err}_{\text{max}}$ is shown in \fig \ref{sfig:norm}. We observe a logarithmic dependence on the requested error bound.

\section{Summary and Outlook}
\label{sec:summary}
In this paper, we present the software package \textit{TimeEvolver}. Its purpose it to compute the exponential of a large sparse matrix that is multiplied with a vector. We specialize to the case in which the matrix is anti-Hermitian. While this may seem like a peculiar choice from a mathematical point of view, it is of utter importance for physics: Any calculation of time evolution can be reduced to this task of exponentiating an anti-Hermitian matrix. In this case, the vector represents the initial state and the matrix is the imaginary unit $i$ times the Hamiltonian. 

It is well-known that an efficient way to tackle this task consists in Krylov subspace methods, and also \textit{TimeEvolver} relies on them. In doing so, \textit{TimeEvolver} goes beyond existing software packages, such as \textit{Expokit} \cite{Sidje1998}, in two ways. First, the specialization to anti-Hermitian matrices makes it possible to incorporate in the numerical implementation a rigorous formula for bounding the error of the Krylov approximation. Moreover, we provide an estimate for additional uncertainties due to numerical roundoff. 
An improved statement about the accuracy of the Krylov approach is of great importance for the potential of \textit{TimeEvolver} to make new discoveries. If our software leads to novel and potentially unexpected results (as was the case in \cite{Dvali2020}), one needs to make sure that they are not artifacts of the numerical method. Since \textit{TimeEvolver} is most useful for systems and phenomena that cannot be studied with any other means, the validity of new discoveries can only be established via an error analysis provided by the program itself.

The second advantage of \textit{TimeEvolver} consists in its ease of use for applications to physics. In order to achieve this, we provide routines to create a basis of possible states and to derive the Hamiltonian matrix from a more abstract representation of the Hamiltonian. In doing so, we specialized to the widely-used case of Hamiltonians that are defined in terms of creation and annihilation operators, but it is straightforward to adapt our program to other operators. Moreover, we have included a concrete physical example, documented all parts of the software package and devised a streamlined installation and build procedure. Finally, our program is open-access and based on free software. In this way, we hope that \textit{TimeEvolver} can contribute to progress across various disciplines in physics.

As an outlook, we would like to point out ideas for future improvements of \textit{TimeEvolver}. First, it is known that a decrease of runtime for Krylov subspace methods can be achieved by employing GPU computing (see \cite{Velamparambil2008} for an early implementation using NVIDIA CUDA\textsuperscript{\circledR}). Secondly, parallelization is capable of bypassing the limitation on the size of the Hilbert space, which arises from the requirement of storing the Hamiltonian matrix in a single memory, by distributing it among many computing units. On a supercomputer, this makes it possible to study Hilbert spaces with dimension close to $10^{10}$ \cite{Brenes2019}. It would be very interesting to combine a parallel implementation of Krylov subspace techniques, such as the one in \cite{Brenes2019}, with the error analysis of \textit{TimeEvolver}.\footnote{Numerous approximate solution techniques have been developed for the study of Hilbert spaces with even larger sizes. Also in such a situation, Krylov subspace techniques can be useful to compute time evolution (see \eg \cite{Paeckel2019} for a review.)}

\section*{Declaration of Competing Interests}
The authors declare that they have no known competing financial interests or personal relationships that could have appeared to influence the work reported in this paper.

\section*{Acknowledgments}
We are grateful to Lukas Eisemann for collaboration in the initial stages of this project, and we thank Gia Dvali for useful comments. Moreover, we are indebted to an anonymous referee for insightful feedback and many helpful suggestions. This work was supported in part, by the Deutsche Forschungsgemeinschaft (DFG, German Research Foundation) under Germany's Excellence Strategy via the Munich Center for Quantum Science and Technology (EXC-2111 - 390814868) and the Excellence Cluster Origins (EXC-2094 - 390783311). 
The work of M.M.\ was supported by a Minerva Fellowship of the Minerva Stiftung Gesellschaft für die Forschung mbH, the Israel Science Foundation (grant No. 741/20) and by the Deutsche Forschungsgemeinschaft through a German-Israeli Project Cooperation (DIP) grant "Holography and the Swampland". The work of S.Z.\ was supported by ERC-AdG-2015 grant 694896.

\bibliographystyle{elsarticle-num}
\bibliography{RefsTimeEvolver}

\begin{thebibliography}{10}
\expandafter\ifx\csname url\endcsname\relax
  \def\url#1{\texttt{#1}}\fi
\expandafter\ifx\csname urlprefix\endcsname\relax\def\urlprefix{URL }\fi
\expandafter\ifx\csname href\endcsname\relax
  \def\href#1#2{#2} \def\path#1{#1}\fi

\bibitem{Sidje1998}
R.~B. Sidje, {Expokit: A Software Package for Computing Matrix Exponentials},
  ACM Trans. Math. Softw. 24~(1) (1998) 130–156.
\newblock \href {http://dx.doi.org/10.1145/285861.285868}
  {\path{doi:10.1145/285861.285868}}.

\bibitem{Park1986}
T.~J. Park, J.~C. Light, {Unitary quantum time evolution by iterative Lanczos
  reduction}, The Journal of Chemical Physics 85~(10) (1986) 5870--5876.
\newblock \href {http://dx.doi.org/10.1063/1.451548}
  {\path{doi:10.1063/1.451548}}.

\bibitem{Gallopoulos1989}
E.~Gallopoulos, Y.~Saad, {On the Parallel Solution of Parabolic Equations}, in:
  Proceedings of the 3rd International Conference on Supercomputing, ICS '89,
  Association for Computing Machinery, New York, NY, USA, 1989, p. 17–28.
\newblock \href {http://dx.doi.org/10.1145/318789.318793}
  {\path{doi:10.1145/318789.318793}}.

\bibitem{Saad1992}
Y.~Saad, {Analysis of Some Krylov Subspace Approximations to the Matrix
  Exponential Operator}, SIAM Journal on Numerical Analysis 29~(1) (1992)
  209--228.
\newblock \href {http://dx.doi.org/10.1137/0729014}
  {\path{doi:10.1137/0729014}}.

\bibitem{Gallopoulos1992}
E.~Gallopoulos, Y.~Saad, {Efficient Solution of Parabolic Equations by Krylov
  Approximation Methods}, SIAM Journal on Scientific and Statistical Computing
  13~(5) (1992) 1236--1264.
\newblock \href {http://dx.doi.org/10.1137/0913071}
  {\path{doi:10.1137/0913071}}.

\bibitem{Sidje1994}
R.~Sidje, {Parallel algorithms for large sparse matrix exponentials:
  Application to numerical transient analysis of Markov processes}, PhD,
  Rennes.

\bibitem{Philippe1995}
B.~Philippe, R.~B. Sidje, {Transient Solutions of Markov Processes by Krylov
  Subspaces}, in: W.~J. Stewart (Ed.), {Computations with Markov Chains},
  Springer US, Boston, MA, 1995, pp. 95--119.
\newblock \href {http://dx.doi.org/10.1007/978-1-4615-2241-6_7}
  {\path{doi:10.1007/978-1-4615-2241-6_7}}.

\bibitem{Druskin1995}
V.~Druskin, L.~Knizhnerman, {Krylov subspace approximation of eigenpairs and
  matrix functions in exact and computer arithmetic}, Numerical Linear Algebra
  with Applications 2~(3) (1995) 205--217.
\newblock \href {http://dx.doi.org/https://doi.org/10.1002/nla.1680020303}
  {\path{doi:https://doi.org/10.1002/nla.1680020303}}.

\bibitem{Stewart1996}
D.~Stewart, T.~Leyk, {Error estimates for Krylov subspace approximations of
  matrix exponentials}, Journal of Computational and Applied Mathematics 72~(2)
  (1996) 359 -- 369.
\newblock \href {http://dx.doi.org/10.1016/0377-0427(96)00006-4}
  {\path{doi:10.1016/0377-0427(96)00006-4}}.

\bibitem{Hochbruck1997}
M.~Hochbruck, C.~Lubich, {On Krylov Subspace Approximations to the Matrix
  Exponential Operator}, SIAM Journal on Numerical Analysis 34~(5) (1997)
  1911--1925.
\newblock \href {http://dx.doi.org/10.1137/S0036142995280572}
  {\path{doi:10.1137/S0036142995280572}}.

\bibitem{Celledoni1997}
E.~Celledoni, I.~Moret, {A Krylov projection method for systems of ODEs},
  Applied Numerical Mathematics 24~(2) (1997) 365 -- 378, second International
  Conference on the Numerical Solution of Volterra and Delay Equations.
\newblock \href {http://dx.doi.org/10.1016/S0168-9274(97)00033-0}
  {\path{doi:10.1016/S0168-9274(97)00033-0}}.

\bibitem{Druskin1998}
V.~Druskin, A.~Greenbaum, L.~Knizhnerman, {Using Nonorthogonal Lanczos Vectors
  in the Computation of Matrix Functions}, SIAM Journal on Scientific Computing
  19~(1) (1998) 38--54.
\newblock \href {http://dx.doi.org/10.1137/S1064827596303661}
  {\path{doi:10.1137/S1064827596303661}}.

\bibitem{Hochbruck1998}
M.~Hochbruck, C.~Lubich, H.~Selhofer, {Exponential Integrators for Large
  Systems of Differential Equations}, SIAM Journal on Scientific Computing
  19~(5) (1998) 1552--1574.
\newblock \href {http://dx.doi.org/10.1137/S1064827595295337}
  {\path{doi:10.1137/S1064827595295337}}.

\bibitem{Lubich2008}
C.~Lubich, {From Quantum to Classical Molecular Dynamics: Reduced Models and
  Numerical Analysis}, European Mathematical Society, 2008.
\newblock \href {http://dx.doi.org/10.4171/067} {\path{doi:10.4171/067}}.

\bibitem{Botchev2013}
M.~A. Botchev, V.~Grimm, M.~Hochbruck, {Residual, Restarting, and Richardson
  Iteration for the Matrix Exponential}, SIAM Journal on Scientific Computing
  35~(3) (2013) A1376--A1397.
\newblock \href {http://dx.doi.org/10.1137/110820191}
  {\path{doi:10.1137/110820191}}.

\bibitem{Ye2013}
Q.~Ye, {Error Bounds for the Lanczos Methods for Approximating Matrix
  Exponentials}, SIAM Journal on Numerical Analysis 51~(1) (2013) 68--87.
\newblock \href {http://dx.doi.org/10.1137/11085935X}
  {\path{doi:10.1137/11085935X}}.

\bibitem{Jia2015}
Z.~Jia, H.~Lv, {A posteriori error estimates of Krylov subspace approximations
  to matrix functions}, Numerical Algorithms 69~(1) (2015) 1--28.
\newblock \href {http://arxiv.org/abs/1307.7219} {\path{arXiv:1307.7219}},
  \href {http://dx.doi.org/10.1007/s11075-014-9878-0}
  {\path{doi:10.1007/s11075-014-9878-0}}.

\bibitem{Wang2016}
H.~Wang, Q.~Ye, {Error Bounds for the Krylov Subspace Methods for Computations
  of Matrix Exponentials}, SIAM Journal on Matrix Analysis and Applications
  38~(1) (2017) 155--187.
\newblock \href {http://arxiv.org/abs/1603.07358} {\path{arXiv:1603.07358}},
  \href {http://dx.doi.org/10.1137/16M1063733} {\path{doi:10.1137/16M1063733}}.

\bibitem{Jawecki2018}
T.~Jawecki, W.~Auzinger, O.~Koch, {Computable upper error bounds for Krylov
  approximations to matrix exponentials and associated $\varphi$-functions},
  BIT Numerical Mathematics 60 (2020) 157–197.
\newblock \href {http://arxiv.org/abs/1809.03369} {\path{arXiv:1809.03369}},
  \href {http://dx.doi.org/10.1007/s10543-019-00771-6}
  {\path{doi:10.1007/s10543-019-00771-6}}.

\bibitem{Saad2003}
Y.~Saad, {Iterative Methods for Sparse Linear Systems}, 2nd Edition, Society
  for Industrial and Applied Mathematics, 2003.
\newblock \href {http://dx.doi.org/10.1137/1.9780898718003}
  {\path{doi:10.1137/1.9780898718003}}.

\bibitem{Parlett1998}
B.~N. Parlett, {The Symmetric Eigenvalue Problem}, Society for Industrial and
  Applied Mathematics, 1998.
\newblock \href {http://dx.doi.org/10.1137/1.9781611971163}
  {\path{doi:10.1137/1.9781611971163}}.

\bibitem{Higham2002}
N.~J. Higham, {Accuracy and Stability of Numerical Algorithms}, 2nd Edition,
  Society for Industrial and Applied Mathematics, 2002.
\newblock \href {http://dx.doi.org/10.1137/1.9780898718027}
  {\path{doi:10.1137/1.9780898718027}}.

\bibitem{takahasi1974double}
H.~Takahasi, M.~Mori, {Double Exponential Formulas for Numerical Integration},
  Publications of the Research Institute for Mathematical Sciences 9~(3) (1974)
  721--741.
\newblock \href {http://dx.doi.org/10.2977/prims/1195192451}
  {\path{doi:10.2977/prims/1195192451}}.

\bibitem{Davis1984}
P.~J. Davis, P.~Rabinowitz, {Methods of Numerical Integration}, 2nd Edition,
  Academic Press, 1984.
\newblock \href {http://dx.doi.org/https://doi.org/10.1016/C2013-0-10566-1}
  {\path{doi:https://doi.org/10.1016/C2013-0-10566-1}}.

\bibitem{Mori2001}
M.~Mori, M.~Sugihara, {The double-exponential transformation in numerical
  analysis}, Journal of Computational and Applied Mathematics 127~(1) (2001)
  287--296.
\newblock \href {http://dx.doi.org/10.1016/S0377-0427(00)00501-X}
  {\path{doi:10.1016/S0377-0427(00)00501-X}}.

\bibitem{Bailey2005}
D.~H. Bailey, K.~Jeyabalan, X.~S. Li, A comparison of three high-precision
  quadrature schemes, Experimental Mathematics 14~(3) (2005) 317--329.
\newblock \href {http://dx.doi.org/10.1080/10586458.2005.10128931}
  {\path{doi:10.1080/10586458.2005.10128931}}.

\bibitem{Kaniel1966}
S.~Kaniel, {Estimates for some computational techniques in linear algebra},
  Math. Comp. 20 (1966) 369--378.
\newblock \href {http://dx.doi.org/10.1090/S0025-5718-1966-0234618-4}
  {\path{doi:10.1090/S0025-5718-1966-0234618-4}}.

\bibitem{Paige1972}
C.~C. Paige, {Computational Variants of the Lanczos Method for the
  Eigenproblem}, IMA Journal of Applied Mathematics 10~(3) (1972) 373--381.
\newblock \href {http://dx.doi.org/10.1093/imamat/10.3.373}
  {\path{doi:10.1093/imamat/10.3.373}}.

\bibitem{Paige1976}
C.~C. Paige, {Error Analysis of the Lanczos Algorithm for Tridiagonalizing a
  Symmetric Matrix}, IMA Journal of Applied Mathematics 18~(3) (1976) 341--349.
\newblock \href {http://dx.doi.org/10.1093/imamat/18.3.341}
  {\path{doi:10.1093/imamat/18.3.341}}.

\bibitem{Paige1980}
C.~Paige, {Accuracy and effectiveness of the Lanczos algorithm for the
  symmetric eigenproblem}, Linear Algebra and its Applications 34 (1980)
  235--258.
\newblock \href {http://dx.doi.org/10.1016/0024-3795(80)90167-6}
  {\path{doi:10.1016/0024-3795(80)90167-6}}.

\bibitem{Dvali2020}
G.~Dvali, L.~Eisemann, M.~Michel, S.~Zell, {Black hole metamorphosis and
  stabilization by memory burden}, Phys. Rev. D 102~(10) (2020) 103523.
\newblock \href {http://arxiv.org/abs/2006.00011} {\path{arXiv:2006.00011}},
  \href {http://dx.doi.org/10.1103/PhysRevD.102.103523}
  {\path{doi:10.1103/PhysRevD.102.103523}}.

\bibitem{Dvali:2012en}
G.~Dvali, C.~Gomez, {Black Holes as Critical Point of Quantum Phase
  Transition}, Eur. Phys. J. C 74 (2014) 2752.
\newblock \href {http://arxiv.org/abs/1207.4059} {\path{arXiv:1207.4059}},
  \href {http://dx.doi.org/10.1140/epjc/s10052-014-2752-3}
  {\path{doi:10.1140/epjc/s10052-014-2752-3}}.

\bibitem{Bekenstein:1973ur}
J.~D. Bekenstein, {Black holes and entropy}, Phys. Rev. D 7 (1973) 2333--2346.
\newblock \href {http://dx.doi.org/10.1103/PhysRevD.7.2333}
  {\path{doi:10.1103/PhysRevD.7.2333}}.

\bibitem{Dvali:2017ktv}
G.~Dvali, {Critically excited states with enhanced memory and pattern
  recognition capacities in quantum brain networks: Lesson from black holes
  }\href {http://arxiv.org/abs/1711.09079} {\path{arXiv:1711.09079}}.

\bibitem{Dvali:2018tqi}
G.~Dvali, M.~Michel, S.~Zell, {Finding Critical States of Enhanced Memory
  Capacity in Attractive Cold Bosons}, EPJ Quant. Technol. 6 (2019) 1.
\newblock \href {http://arxiv.org/abs/1805.10292} {\path{arXiv:1805.10292}},
  \href {http://dx.doi.org/10.1140/epjqt/s40507-019-0071-1}
  {\path{doi:10.1140/epjqt/s40507-019-0071-1}}.

\bibitem{Dvali:2018xpy}
G.~Dvali, {A Microscopic Model of Holography: Survival by the Burden of Memory
  }\href {http://arxiv.org/abs/1810.02336} {\path{arXiv:1810.02336}}.

\bibitem{Hawking:1974sw}
S.~W. Hawking, {Particle Creation by Black Holes}, Commun. Math. Phys. 43
  (1975) 199--220, [Erratum: Commun. Math. Phys. 46, 206 (1976)].
\newblock \href {http://dx.doi.org/10.1007/BF02345020}
  {\path{doi:10.1007/BF02345020}}.

\bibitem{Velamparambil2008}
S.~Velamparambil, S.~MacKinnon-Cormier, J.~Perry, R.~Lemos, M.~Okoniewski,
  J.~Leon, {GPU Accelerated Krylov Subspace Methods for Computational
  Electromagnetics}, in: {2008 38th European Microwave Conference}, 2008, pp.
  1312--1314.
\newblock \href {http://dx.doi.org/10.1109/EUMC.2008.4751704}
  {\path{doi:10.1109/EUMC.2008.4751704}}.

\bibitem{Brenes2019}
M.~Brenes, V.~Varma, A.~Scardicchio, I.~Girotto, {Massively parallel
  implementation and approaches to simulate quantum dynamics using Krylov
  subspace techniques}, Computer Physics Communications 235 (2019) 477--488.
\newblock \href {http://arxiv.org/abs/1704.02770} {\path{arXiv:1704.02770}},
  \href {http://dx.doi.org/10.1016/j.cpc.2018.08.010}
  {\path{doi:10.1016/j.cpc.2018.08.010}}.

\bibitem{Paeckel2019}
S.~Paeckel, T.~Köhler, A.~Swoboda, S.~R. Manmana, U.~Schollwöck, C.~Hubig,
  {Time-evolution methods for matrix-product states}, Annals of Physics 411
  (2019) 167998.
\newblock \href {http://arxiv.org/abs/1901.05824} {\path{arXiv:1901.05824}},
  \href {http://dx.doi.org/https://doi.org/10.1016/j.aop.2019.167998}
  {\path{doi:https://doi.org/10.1016/j.aop.2019.167998}}.

\end{thebibliography}

\end{document}